\documentclass[prc,notitlepage, twocolumn,
showpacs,floatfix,nofootinbib,preprintnumbers,superscriptaddress,amsmath,amssymb]{revtex4-1}

\usepackage{graphicx}\usepackage{epsfig}
\usepackage{amsmath,amssymb}
\usepackage{bm}
\usepackage{color}
\usepackage{float}
\usepackage{dcolumn} 
\usepackage{enumerate}
\usepackage{multirow}
\usepackage{threeparttable}

\begin{document}

\newcommand{\Ct}{\tilde{C}}
\newcommand{\rhot}{\tilde{\rho}}
\newcommand{\taut}{\tilde{\tau}}
\newcommand{\Jt}{\tilde{J}}
\newcommand{\Jbt}{\tilde{\bm{J}}}
\newcommand{\Jsft}{\tilde{\sf J}}
\newcommand{\sbt}{\tilde{\bm{s}}}
\newcommand{\Jb}{\bm{J}}
\newcommand{\rb}{\bm{r}}
\newcommand{\rbp}{\bm{r}'}
\newcommand{\pb}{\bm{p}}

\title{
  Energy-weighted sum rule for nuclear density functional theory
  }
\author{Nobuo Hinohara}
\affiliation{%
  Center for Computational Sciences, University of Tsukuba, Tsukuba 305-8577, Japan
}
\affiliation{
  Faculty of Pure and Applied Sciences, University of Tsukuba, Tsukuba 305-8571, Japan
}

\date{\today}

\begin{abstract}
  The expressions for the energy-weighted sum rule of the isoscalar and isovector coordinate operators are derived based on the second-order fluctuation of the local densities.
  Conventional derivation of the Thouless theorem for the energy-weighted sum rule is based on the double commutator of the Hamiltonian, while the present derivation does not assume a Hamiltonian operator and is applicable to nuclear energy density functionals.
The expressions include the contribution of the local gauge symmetry breaking of the energy density functional.
It is shown that the local gauge invariance of the kinetic and current densities and kinetic pair density is important, while
  all the other local densities do not contribute to the energy-weighted sum rule of the coordinate operators.  
  The finite-amplitude method calculations are performed and the expressions for the energy-weighted sum rule are numerically examined for the isoscalar and isovector multipole operators up to $L=3$ for selected spherical and axially deformed nuclei.
\end{abstract}


\maketitle

\section{Introduction}

In atomic nuclei 
there are numerous excited states that originate from the
single-particle and collective motion of the constituent nucleons.
Thus it is useful to have a few representative quantities of the excited states. 
The sum rule \cite{Bohigas1979267,Lipparini1989103} is a quantity which involves all the excited states, and contains important collective information on the properties of the excited states, such as the giant resonances \cite{Harakeh-Woude} and
the Nambu-Goldstone modes \cite{PhysRevC.92.034321,PhysRevC.97.034321}.

The energy-weighted sum rule is the most commonly used one among various energy moments of the sum rules.
Although it is a summation over all the excited states, 
the Thouless theorem \cite{Thouless196178} allows one to evaluate a sum-rule value that is 
the summation over all the excited states computed through the
random-phase approximation (RPA), using the expectation value of the double commutator of the Hamiltonian at the ground state computed within the self-consistent Hartree-Fock (HF) theory.
The theorem has been proved also for the Hartree-Fock-Bogoliubov (HFB) + quasiparticle RPA (QRPA)
\cite{PhysRevC.66.024309} and the second RPA \cite{PhysRevC.90.024305,PhysRevC.35.1159}.
The double commutator of the Hamiltonian becomes simple for the isoscalar and isovector coordinate operators. In the zero-range Skyrme force, only the kinetic-energy term in the Hamiltonian contributes to the energy-weighed sum for an isoscalar coordinate operator, and the kinetic-energy term and momentum-dependent terms in the interaction contribute to the energy-weighted sum rule of an isovector coordinate operator.
Therefore, the Thouless theorem significantly reduces the computational costs of the energy-weighted sum rule, and is also useful for verifying the accuracy of the QRPA calculation.

Nuclear density functional theory (DFT) can be regarded as a starting point of the mean-field models \cite{RevModPhys.75.121,RevModPhys.88.045004}.
In nuclear DFT, the form of the energy density functional (EDF) is not given a priori.
Several EDFs based on the nonrelativistic Skyrme and Gogny forces and on relativistic theory are widely used.
The EDF of these types can be derived from the corresponding effective interaction. In that case one can go back to the Hamiltonian (effective interaction) starting from the EDF.
However, in general, there is no direct correspondence 
to the effective interaction in the nuclear DFT,
if the EDF and its coupling constants are constructed directly
by reproducing a representative set of the experimental observables.
The existence of the Hamiltonian operator is not guaranteed.

Although the Thouless theorem has been applied widely within the framework of the nuclear DFT, to the best of my knowledge
it has not been proved for the nuclear DFT where the EDF does not correspond to a Hamiltonian operator, and thus
the double-commutator expression cannot be justified.
This includes the case when
the EDF is constructed independently of the interactions (such as UNEDF functionals \cite{PhysRevC.82.024313,PhysRevC.85.024304,PhysRevC.89.054314,0954-3899-42-3-034024}).
Even the standard Skyrme HFB calculation is not carried out within the two-body and three-body Skyrme effective interaction.
Prescriptions used in the spin-orbit and tensor functional may break the correspondence with the Hamiltonian.
The Skyrme spin-orbit interaction has a single interaction strength $W_0$ and it determines the
isoscalar and isovector coupling constants of the spin-orbit functional.
In several Skyrme EDFs, an additional parameter $b_4'$ is introduced to control the isovector property
of the spin-orbit functional \cite{Reinhard1995467}.
The tensor-density (spin-current density) terms appear from the momentum-dependent $t_1$ and $t_2$ terms of the Skyrme effective interaction even without including the tensor effective interactions ($t_e$ and $t_o$ terms).
However, because of the complicated treatment of the tensor-density terms in deformed nuclei, the contribution from this term is often neglected except for a few parameter sets such as SLy5 \cite{Chabanat1998231} and SkP \cite{Dobaczewski1984103}.
Moreover, the connection to the Hamiltonian operator is lost by the existence of the two-body density-dependent term (however, it has been shown that the density-dependent force does not contribute in the Thouless theorem \cite{Bohigas1979267}). 
Another issue is the treatment of the pairing interaction.
Except for the SkP interaction, the pairing interaction used in the standard Skyrme HFB calculation has a simple form and density dependence, and is independent of the particle-hole interaction, while in the mean-field approach starting from an effective interaction, the same interaction should provide the Hartree-Fock potential and pairing potential.

In a previous work \cite{PhysRevC.91.044323},
it was numerically shown that in the SLy4 EDF 
the inclusion of the time-odd current terms is necessary to recover the energy-weighted sum-rule values of the Thouless theorem,
and that other terms in the time-odd functional do not impact the values of the energy-weighted sum rule at all.
The time-odd current terms are necessary in order to satisfy the Galilean invariance of the EDF.
More generalized forms of the EDF could be used in the future, and thus
it is desired to understand the applicability of the Thouless theorem to the nuclear DFT.
Note that Kerman-Onishi condition can be derived for the nuclear EDF
from the transformation of the densities without assuming the Hamiltonian operator \cite{PhysRevC.88.034311,PhysRevC.84.064303},
and that the lack of a relation with the Hamiltonian formalism can cause
problems when evaluating the energy of the quantum-number projected state within the nuclear DFT
\cite{PhysRevC.79.044318,PhysRevC.79.044318,PhysRevC.79.044319,PhysRevC.79.044320}.

The aim of this paper is to derive the expression for the energy-weighted sum rule
within the nuclear DFT without using the double commutator of the Hamiltonian.
By considering a fluctuation to the HFB state, and comparing the fluctuation of the energy in two ways, the expression of the energy-weighted sum rule is derived.
This derivation can be applied to the nuclear EDF which does not have a corresponding Hamiltonian operator.

This paper is organized as follows.
In Sec.~\ref{sec:EDF},
the nuclear EDF is introduced.
Section~\ref{sec:Thouless}
recapitulates the conventional derivation of the Thouless theorem based on the double commutator of the Hamiltonian, then presents the derivation for the nuclear EDF.
Section~\ref{sec:fam} summarizes the energy-weighted sum-rule calculation based on the complex-energy finite-amplitude method.
In Sec.~\ref{sec:sumrule}, energy-weighted sum-rule values of various multipole operators are numerically calculated using the complex-energy finite-amplitude method, and are compared with the values of the Thouless theorem derived for general nuclear EDFs.
Conclusions are given in Sec.~\ref{sec:conclusion}.

\section{Nuclear EDF \label{sec:EDF}}

I consider a general form of the nuclear EDF of Skyrme type
that is quadratic in local densities (except for the density-dependent terms)
and can contain up to two spacial derivatives
but without neutron-proton mixing \cite{ActaPhysPolB27_45,PhysRevC.69.014316}.
The nuclear EDF has the following form:
\begin{align}
  E[\rho,\tilde{\rho}] &= \int d\rb {\cal E}(\rb),\\
  {\cal E}(\rb) &= \frac{\hbar^2}{2m}\tau_0(\rb) + \sum_{k=0}^1 \chi_k(\rb) + {\cal E}_{\rm Coul}(\rb) +
  \sum_{t=n,p} \tilde{\chi}_t (\rb),  \label{eq:EDF}
\end{align}
where the first term in Eq.~(\ref{eq:EDF}) is the isoscalar kinetic energy,
$\chi_k$ are the isoscalar ($k=0$) and isovector ($k=1$) particle-hole EDFs,
${\cal E}_{\rm Coul}$ is the Coulomb EDF, and 
$\tilde{\chi}_t$ are the neutron ($t=n=1/2$) and proton ($t=p=-1/2$) pairing EDFs.
Throughout this paper, I use the index $k$ to specify the isoscalar or isovector character, and the index $t$ for neutrons or protons.

The particle-hole EDF is given by its time-even and time-odd parts,
\begin{align}
  \chi_k(\rb) &= \chi_k^{\rm even}(\rb) + \chi_k^{\rm odd}(\rb), \\
  \chi_k^{\rm even}(\rb) &=
  C^\rho_k[\rho_0] \rho_k^2  + C^{\Delta\rho}_k \rho_k \Delta\rho_k
  + C^\tau_k \rho_k\tau_k + C_k^{J0} J_k^2\nonumber  \label{eq:phEDF} \\
  &\quad 
    + C_k^{J1} \bm{J}_k^2
    + C_k^{J2}\underline{\sf J}_k^2 
    + C^{\nabla J}_k \rho_k\bm{\nabla}\cdot \bm{J}_k, \\
  \chi_k^{\rm odd}(\rb) &=
    C_k^s[\rho_0] \bm{s}_k^2 + C_k^{\Delta s} \bm{s}_k \cdot \Delta \bm{s}_k
  + C_k^T \bm{s}_k\cdot\bm{T}_k  
  + C_k^j \bm{j}_k^2 \nonumber \\
 &\quad + C_k^{\nabla j} \bm{s}_k \cdot (\bm{\nabla}\times \bm{j}_k)
  + C_k^{\nabla s} (\bm{\nabla}\cdot \bm{s}_k)^2 \nonumber \\
 &\quad + C_k^F \bm{s}_k \cdot \bm{F}_k. \label{eq:ppEDF}
\end{align}
The time-even part is composed of the particle-hole density $\rho_k$, kinetic density $\tau_k$, and pseudoscalar, pseudovector, and pseudotensor densities $J_k$, $\bm{J}_k$, and $\underline{\sf J}_k$.
The time-odd parts are described with the spin density $\bm{s}_k$, spin-kinetic density $\bm{T}_k$, current density $\bm{j}_k$, and tensor-kinetic density $\bm{F}_k$. Definitions of these local densities are summarized in Appendix \ref{sec:densities}.
Some of the coupling constants $C^\rho_k$ and $C^s_k$ have isoscalar particle-hole density dependence ($C_k[\rho_0] = C_{k0} + C_{k{\rm D}} \rho_0^\gamma$).
In the Skyrme force, all the coupling constants are basically derived from the effective interactions, while in the UNEDF optimizations
\cite{PhysRevC.82.024313,PhysRevC.85.024304,PhysRevC.89.054314,0954-3899-42-3-034024}
only the time-even coupling constants are optimized using experimental data.
For the even-even systems with time-reversal symmetry, the time-odd functionals turn on only in the linear response calculation.
The Coulomb functional is composed of direct and exchange terms, which are functionals of the proton particle-hole density only [$\rho_p=(\rho_0 - \rho_1)/2$]:
\begin{align}
  {\cal E}_{\rm Coul}(\rb) &= {\cal E}_{\rm dir}(\rb) + {\cal E}_{\rm ex}(\rb), \\
  {\cal E}_{\rm dir}(\rb) &= \frac{1}{2}e^2 \rho_p(\rb) \int d\rbp \frac{\rho_p(\rbp)}{|\rb-\rbp|}, \\
  {\cal E}_{\rm ex}(\rb) &=  -e^2 \frac{3}{4}\left(\frac{3}{\pi}\right)^{\frac{1}{3}} \rho_p(\rb)^{\frac{4}{3}}.
\end{align}
The general form of the pairing EDF that is quadratic in local pair densities is given by
\begin{align}
  \tilde{\chi}_t(\rb) &= \Ct^\rho_t[\rho_0] |\rhot_t|^2
  + \Ct^{\Delta\rho}_t {\rm Re}( \rhot^\ast_t \Delta\rhot_t)
  + \Ct^\tau_t {\rm Re}(\rhot^\ast_t \taut_t) \nonumber \\
  &\quad + \Ct^{J0}_t |\Jt_t|^2
  + \Ct^{J1}_t |\Jbt_t|^2
  + \Ct^{J2}_t |\underline{\Jsft}_t|^2 \nonumber \\
  &\quad + \Ct^{\nabla J}_t {\rm Re} ( \rhot^\ast_t \bm{\nabla}\cdot \Jbt_t)
\end{align}
with the pair density $\rhot_t$, kinetic pair density $\taut_t$, and tensor pair densities
$\Jt_t$, $\Jbt_t$, and $\underline{\Jsft}_t$.
In most of the Skyrme EDFs, only the first term with an isoscalar particle-hole density dependence is used in the pairing EDF:
\begin{align}
  \Ct^\rho_t[\rho_0] = \frac{V_t}{4}\left( 1 - \eta_t \frac{\rho_0(\rb)}{\rho_c}\right), \label{eq:pairforce}
\end{align}
where $V_t$ is the strength and $\eta_t$ controls the isoscalar particle-hole density dependence.

\section{Thouless theorem for energy-weighted sum rule \label{sec:Thouless}}

\subsection{Operator derivation}

First I recapitulate the conventional derivation of the Thouless theorem \cite{Thouless196178}
based on the discussion in Ref.~\cite{Bohigas1979267}.
I consider a system described by a Hamiltonian of the Skyrme interaction:
\begin{align}
  \hat{H} &= \hat{T}+\hat{V},\\
  \hat{T} &= \frac{1}{2m} \sum_{i=1}^A \hat{\bm{p}}_i^2,\\
  \hat{V} &= \sum_{i<j}
  t_0 (1 + x_0 \hat{P}^\sigma) \delta(\hat{\rb}_{ij}) \nonumber\\
  &\quad + \frac{t_1}{2}
   ( 1 + x_1\hat{P}^\sigma) [\hat{\bm{k}}^{\prime 2}\delta(\hat{\rb}_{ij}) + \delta(\hat{\rb}_{ij})\hat{\bm{k}}^2] \nonumber \\
  &\quad + t_2 (1 + x_2 \hat{P}^\sigma )\hat{\bm{k}}'\cdot \delta(\hat{\rb}_{ij})\hat{\bm{k}} \nonumber \\
  &\quad + \frac{t_3}{6}( 1+x_3\hat{P}^\sigma) \rho^\gamma\left(\frac{\rb_1+\rb_2}{2}\right)\delta(\hat{\rb}_{ij}) \nonumber \\
  &\quad + \frac{t_e}{2}[ \hat{\bm{k}}^{\prime}\cdot \hat{\sf S}\cdot\hat{\bm{k}}^{\prime} \delta(\hat{\rb}_{ij}) + \delta(\hat{\rb}_{ij}) \hat{\bm{k}}\cdot\hat{\sf S}\cdot\hat{\bm{k}}] \nonumber \\ 
  &\quad + t_o \hat{\bm{k}}^{\prime}\cdot \hat{\sf S} \delta(\hat{\rb}_{ij})\cdot \hat{\bm{k}}\nonumber \\
  &\quad + i W_0 (\hat{\bm{\sigma}}_i + \hat{\bm{\sigma}}_j)\cdot [ \hat{\bm{k}}^{\prime} \times  \delta(\hat{\rb}_{ij})\hat{\bm{k}}],
\end{align}
where $\hat{\rb}_{ij}=\hat{\rb}_i-\hat{\rb}_j$, 
$\hat{P}^\sigma=(1+\hat{\bm{\sigma}}_i\cdot\hat{\bm{\sigma}}_j)/2$ is the spin-exchange operator,
$\hat{\sf S}=3(\hat{\bm{\sigma}}_i\cdot\bm{e}_r)(\hat{\bm{\sigma}}_j\cdot\bm{e}_r) - \hat{\bm{\sigma}}_i\cdot\hat{\bm{\sigma}}_j$ is the tensor operator, and 
\begin{align}
  \hat{\bm{k}}  &= \frac{1}{2i}( \bm{\nabla}_i - \bm{\nabla}_j), \\
  \hat{\bm{k}}' &= -\frac{1}{2i}( \bm{\nabla}_i - \bm{\nabla}_j).
\end{align}
The energy-weighted sum rule of an operator $\hat{F}$ is expressed in terms of the double commutator of the Hamiltonian:
\begin{align}
  m_1(\hat{F}) &=
   \sum_{\lambda, \Omega_\lambda>0} \Omega_\lambda | \langle \lambda|\hat{F}|0\rangle|^2  \nonumber \\
  &= -\frac{1}{2}\langle\Psi_{\rm HFB}|
  \bm{[}[ \hat{H}, \hat{F}], \hat{F}\bm{]}|\Psi_{\rm HFB}\rangle,  \label{eq:m1}
\end{align}
where $|\Psi_{\rm HFB}\rangle$ is the HFB state, $|0\rangle$ is the QRPA correlated ground state, and $|\lambda\rangle$ is the QRPA $\lambda$th excited state with an excitation energy $\Omega_\lambda=E_\lambda - E_0$.
When the operator $\hat{F}$ is an isoscalar-coordinate type,
\begin{align}
  \hat{F}^{\rm IS}=\alpha \sum_{i=1}^A f(\hat{\rb}_i),
  \end{align}
it can be shown that the double commutator of the interaction term cancels,
and the contribution to the energy-weighted sum rule is from the momentum operator in the kinetic-energy term in the Skyrme interaction:
\begin{align}
m_1(\hat{F}^{\rm IS})&=
  -\frac{1}{2} \langle \bm{[}[\hat{T}, \hat{F}^{\rm IS}], \hat{F}^{\rm IS}\bm{]}\rangle 
  = \alpha^2 \frac{\hbar^2}{2m} \sum_{i=1}^A \langle [\nabla f(\hat{\rb}_i)]^2\rangle \nonumber \\
  &= \alpha^2 \frac{\hbar^2}{2m} \int d\rb [\nabla f(\rb)]^2\rho_0(\rb).
  \label{eq:isewsr}
\end{align}
The momentum-independent terms with $t_0$ and $t_3$ are shown to commute with the coordinate operator.
The $t_1$ and $t_2$ terms can be written as
\begin{align}
  \hat{V}_{t_1,t_2}&= \frac{1}{2} \sum_{i,j=1}^A \left(
  \frac{t_1}{8\hbar^2} \{\hat{\bm{p}}_{ij}^2, \delta(\hat{\rb}_{ij})\} + \frac{t_2}{4\hbar^2} \hat{\bm{p}}_{ij}\delta(\hat{\rb}_{ij}) \hat{\bm{p}}_{ij}
  \right) \nonumber \\
  &= \frac{1}{8\hbar^2}  \sum_{i,j=1}^A \biggl\{
  \frac{t_1}{2}\bm{[} \hat{\bm{p}}_{ij}, [ \hat{\bm{p}}_{ij}, \delta(\hat{\rb}_{ij})]\bm{]}
  \nonumber \\
  &\quad + (t_1+t_2) \hat{\bm{p}}_{ij} \delta(\hat{\rb}_{ij}) \hat{\bm{p}}_{ij} \biggr\}, \label{eq:Vpot}
\end{align}
where $\hat{\pb}_{ij}=\hat{\pb}_i - \hat{\pb}_j$.
The first term is the second derivative of the $\delta$ function, and it commutes with any coordinate operators.
The commutator with the second term is shown to be 
\begin{align}
  [\hat{V}_{t_1,t_2},\hat{F}^{\rm IS}] &=
  \frac{1}{8\hbar^2} (t_1+t_2) \sum_{i,j,k=1}^A [ \hat{\bm{p}}_{ij}\delta(\hat{\rb}_{ij})\hat{\bm{p}}_{ij}, f(\hat{\rb}_k)] \nonumber \\
  &= \frac{t_1+t_2}{8\hbar^2} \sum_{i,j,k=1}^A  \{  \hat{\bm{p}}_{ij}, [\hat{\bm{p}}_{ij}, f(\hat{\rb}_k)]\delta(\hat{\rb}_{ij}) \} \nonumber \\
  &= -i\frac{t_1+t_2}{4\hbar}\sum_{i,j=1}^A\left\{ \hat{\bm{p}}_{ij}, [\bm{\nabla}f(\hat{\rb}_i)] \delta(\hat{\rb}_{ij})\right\}\nonumber\\
  &= 0,
\end{align}
as interchanging $i$ and $j$ changes the sign.
In the same way, one can derive that the commutators with the $t_e$, $t_o$, and $W_0$ terms become zero.

For the isovector operator
\begin{align}
\hat{F}^{\rm IV}=\sum_{i=1}^A \alpha_{t_i} f(\hat{\rb}_i) \tau^1(t_i), \label{eq:iv}
\end{align}
where $\tau^1(t_i) = 2t_i$,
generally both the kinetic and interaction parts of the Hamiltonian
contribute to the energy-weighted sum rule \cite{Lipparini1989103},
\begin{align}
  m_1(\hat{F}^{\rm IV}) &= -\frac{1}{2}\langle \bm{[}[ \hat{T}+\hat{V}, \hat{F}^{\rm IV}], \hat{F}^{\rm IV}\bm{]} \rangle \nonumber\\
  &= m_1^{\rm kin}(\hat{F}^{\rm IV})\left[1 + \kappa(\hat{F}^{\rm IV})\right],
\end{align}
where $m_1^{\rm kin}(\hat{F}^{\rm IV})$ is the contribution from the kinetic energy
\begin{align}
  m_1^{\rm kin}(\hat{F}^{\rm IV}) &= -\frac{1}{2}\langle\bm{[}[ \hat{T},\hat{F}^{\rm IV}], \hat{F}^{\rm IV}\bm{]}\rangle \nonumber \\
  &= \frac{\hbar^2}{2m} \sum_{i=1}^A \alpha_{t_i}^2 \langle [\nabla f(\hat{\rb}_i)]^2\rangle \nonumber \\
  &= \frac{\hbar^2}{2m} \int d\rb [\nabla f(\rb)]^2 [\alpha_n^2\rho_n(\rb) + \alpha_p^2\rho_p(\rb)],
  \label{eq:ivewsr}
\end{align}
and the enhancement factor $\kappa(\hat{F}^{\rm IV})$ shows the relative contribution of the interaction-energy term with respect to the kinetic part to the energy-weighted sum rule.
The potential contribution is from the second term in Eq.~(\ref{eq:Vpot}).
The spin-exchange parts with $x_1$ and $x_2$
also contribute with factor $\frac{1}{2}$ from $\hat{P}^\sigma$ operator, as the $\bm{\sigma}_i\cdot\bm{\sigma}_j$ part  produces the spin density which is zero for even-even systems:
\begin{align}
  m_1^{\rm kin} \kappa(\hat{F}^{\rm IV}) &= -\frac{1}{2}\langle\bm{[}[\hat{V}, \hat{F}^{\rm IV}], \hat{F}^{\rm IV}\bm{]} \rangle
  \nonumber \\
  &=   \frac{t_1(2+x_1)+t_2(2+x_2)}{8} \sum_{i,j=1}^A \alpha_{t_i} \tau^1(t_i)  \nonumber \\
  &\quad \times \biggl\{ \alpha_{t_i} \tau^1(t_i)\langle [\nabla f(\hat{\rb}_i)]^2 \delta(\hat{\rb}_{ij})\rangle \nonumber \\
  &\quad  - \alpha_{t_j} \tau^1(t_j)\langle \bm{\nabla}f(\hat{\rb}_i)\cdot \bm{\nabla}f(\hat{\rb}_j)\delta(\hat{\rb}_{ij})\rangle \biggr\} \nonumber \\
  &=
  \frac{t_1(2+x_1)+t_2(2+x_2)}{8}(\alpha_n+\alpha_p)^2 \nonumber \\
  & \quad \times 
         \displaystyle \int d\rb [\nabla f(\rb)]^2 \rho_n(\rb)\rho_p(\rb), \label{eq:ivkappa}
\end{align}
where $\frac{1}{8}[t_1(2+x_1)+t_2(2+x_2)] = C^\tau_0 - C^\tau_1$.

These expressions for the energy-weighted sum rule are based on the operator expressions of the kinetic and interaction terms.
Strictly speaking, in the case of the nuclear EDF, in which there is
no correspondence between the EDF and the Hamiltonian operator $\hat{H}$, Eqs.~(\ref{eq:isewsr}), (\ref{eq:ivewsr}), and (\ref{eq:ivkappa}) cannot be derived in the same manner.
In the next subsection  the expressions for the energy-weighted sum rule are derived without
assuming the Hamiltonian operator.

\subsection{Derivation for nuclear EDF}

Following the discussion in Sec.~10.2 of Ref.~\cite{Blaizot-Ripka},
I show that the energy-weighted sum rule is expressed as the second-order fluctuation of the total energy. I consider a small fluctuation starting from a HFB state $|\Psi_{\rm HFB}\rangle$.
As the HFB state is a vacuum of quasiparticles, $\hat{a}_\mu|\Psi_{\rm HFB}\rangle=0$,
such a fluctuation from the HFB state can be described by a quasiparticle-quasihole, quasiparticle-quasiparticle, and quasihole-quasihole densities.
The quasihole-quasihole and quasiparticle-quasihole densities are given by
\begin{align}
  \overline{\kappa}_{\mu\nu} &= \langle \Phi'| \hat{a}_\nu \hat{a}_\mu|\Phi'\rangle, \label{eq:qhqhdensity} \\
    \overline{\rho}_{\mu\nu} &= \langle \Phi'| \hat{a}_\nu^\dag \hat{a}_\mu|\Phi'\rangle,
\end{align}
where the state $|\Phi'\rangle$ includes a small fluctuation.
The coherent state representation of the state $|\Phi'\rangle$ gives that $\bar{\rho}$ is higher order in $\bar{\kappa}$, $\bar{\rho}\sim (\bar{\kappa}\bar{\kappa}^\dag)$.
Therefore the small-amplitude expansion of the energy from the HFB state is given as an expansion with respect to $\bar{\kappa}$ and $\bar{\kappa}^\ast$:
\begin{align}
  E'[\bar{\kappa},\bar{\kappa}^\ast]  =
  E_0' + \frac{1}{2}
  \begin{pmatrix} \bar{\kappa}^\ast & \bar{\kappa} \end{pmatrix}
  \begin{pmatrix}  A & B \\ B^\ast & A^\ast \end{pmatrix}
  \begin{pmatrix} \bar{\kappa} \\ \bar{\kappa}^\ast \end{pmatrix}
   + O( |\bar{\kappa}|^3),
\end{align}
where $E_0'$ is the HFB value of the EDF (with particle-number constraint term),
and $A$ and $B$ are the QRPA matrices given by
\begin{align}
  A_{\rho\sigma,\mu\nu} &= \delta_{\rho\mu} \delta_{\sigma\nu} (E_\mu + E_\nu)
  + \frac{\partial^2 E'}{\partial \bar{\kappa}^\ast_{\rho\sigma}
    \partial \bar{\kappa}_{\mu\nu}}, \\
  B_{\rho\sigma,\mu\nu} &= \frac{\partial^2 E'}{\partial \bar{\kappa}^\ast_{\rho\sigma}
    \partial \bar{\kappa}^\ast_{\mu\nu}},
\end{align}
with the quasiparticle energies $E$.

Suppose that this small fluctuation is given with a Hermitian operator $\hat{F}$:
\begin{align} 
  |\Phi'\rangle = e^{i\eta\hat{F}}|\Psi_{\rm HFB}\rangle, \label{eq:perturbation}
\end{align}
where $\eta$ is a small real parameter.
The operator $\hat{F}$ is written in the quasiparticle representation as
\begin{align}
  \hat{F} &= \langle\Psi_{\rm HFB}|\hat{F}|\Psi_{\rm HFB}\rangle 
  + \sum_{\mu<\nu} \left\{ F^{20}_{\mu\nu} \hat{a}^\dag_\mu \hat{a}^\dag_\nu
  + F^{02}_{\mu\nu} \hat{a}_\nu \hat{a}_\mu \right\} \nonumber \\ & \quad
  + \sum_{\mu\nu} F^{11}_{\mu\nu} \hat{a}^\dag_\mu \hat{a}_\nu,
  \label{eq:Fqp}
\end{align}
where $F^{02}=F^{20\ast}$. From Eqs.~(\ref{eq:perturbation}) and (\ref{eq:Fqp})
one can express the quasihole-quasihole densities $\bar{\kappa}$ in Eq. (\ref{eq:qhqhdensity}) in terms of the matrix element $F^{20}$ and $F^{02}$ as
\begin{align}
  \overline{\kappa}_{\mu\nu} &=
  \langle \Psi_{\rm HFB}|e^{-i\eta\hat{F}} \hat{a}_\nu \hat{a}_\mu e^{i\eta\hat{F}}|\Psi_{\rm HFB}\rangle = -i\eta F^{20}_{\mu\nu},\\ 
  \overline{\kappa}^\ast_{\mu\nu} &=
  \langle \Psi_{\rm HFB}|e^{-i\eta\hat{F}} \hat{a}^\dag_\mu \hat{a}^\dag_\nu e^{i\eta\hat{F}}|\Psi_{\rm HFB}\rangle = i\eta F^{02}_{\mu\nu}.
\end{align}
The energy of this state with the fluctuation $|\Phi'\rangle$ is given by
\begin{align}
  E'[ -i\eta F^{20}, i\eta F^{02}]
  = E_0' + \eta^2 m_1(\hat{F}) + O(\eta^3), \label{eq:qpexp}
\end{align}
where
\begin{align}
  m_1(\hat{F}) =  \frac{1}{2} \begin{pmatrix} F^{02} & -F^{20} \end{pmatrix}
  \begin{pmatrix} A & B \\ B^\ast & A^\ast \end{pmatrix}
  \begin{pmatrix} F^{20} \\ -F^{02} \end{pmatrix}. \label{eq:m1fromab}
\end{align}
Equation (\ref{eq:m1fromab}) is derived by applying the QRPA equations
\begin{align}
  \begin{pmatrix} A & B \\ -B^\ast & -A^\ast \end{pmatrix}
  \begin{pmatrix} X^\lambda \\ Y^\lambda \end{pmatrix}
  = \Omega_\lambda
  \begin{pmatrix} X^\lambda \\ Y^\lambda \end{pmatrix}
\end{align}
and the expression for the transition strength
\begin{align}
  \langle \lambda | \hat{F} | 0\rangle = \sum_{\mu<\nu} \left( X^{\lambda\ast}_{\mu\nu} F^{20}_{\mu\nu}+ Y^{\lambda\ast}_{\mu\nu}
  F^{02}_{\mu\nu} \right)
\end{align}
to Eq.~(\ref{eq:m1}) \cite{Ring-Schuck, Blaizot-Ripka, PhysRevC.79.054329}.
Equation (\ref{eq:qpexp}) shows that the energy-weighted sum rule $m_1(\hat{F})$ appears
as a second-order fluctuation of the total energy of the system where the fluctuation is produced by the operator $\hat{F}$
in the form of Eq. (\ref{eq:perturbation}).

When a Hamiltonian operator exists ($\hat{H}'=\hat{H}-\sum_{t=n,p} \lambda_t \hat{N}^t)$,
the energy of the perturbed state $|\Phi'\rangle$ is given by the expectation value of the Hamiltonian,
\begin{align}
  \langle \Phi'| \hat{H}'| \Phi'\rangle
  &= \langle \Psi_{\rm HFB}|e^{-i\eta \hat{F}} \hat{H}' e^{i\eta\hat{F}} |\Psi_{\rm HFB}\rangle\nonumber \\
  &= \langle \Psi_{\rm HFB}| \hat{H}' + i\eta [\hat{H}',\hat{F}]
  - \frac{\eta^2}{2}\bm{[}[ \hat{H}', \hat{F}], \hat{F}\bm{]} \nonumber \\
  &\quad + O(\eta^3) |\Psi_{\rm HFB}\rangle.
\end{align}  
By comparing the term proportional to $\eta^2$ with Eq.~(\ref{eq:qpexp}), one can derive the Thouless theorem
in the double commutator form \cite{Blaizot-Ripka,doi:10.1093/ptep/ptx004}
\begin{align}
  m_1(\hat{F}) &= \left.\frac{1}{2}\frac{\partial^2}{\partial \eta^2}\langle \Psi_{\rm HFB}|e^{-i\eta\hat{F}}\hat{H}'e^{i\eta\hat{F}}|\Psi_{\rm HFB}\rangle\right|_{\eta=0} \nonumber \\ 
&=  - \frac{1}{2}\langle \Psi_{\rm HFB}|  \bm{[}[ \hat{H}', \hat{F}], \hat{F}\bm{]}| \Psi_{\rm HFB}\rangle. \label{eq:EWSR}
\end{align}
In the case of the nuclear EDF, the total energy of the perturbed state, Eq.~(\ref{eq:perturbation}), is
expressed using the densities evaluated with the perturbed state instead of the Hamiltonian operator. Equation~(\ref{eq:qpexp}) can be written as
\begin{align}
  m_1(\hat{F}) = \frac{1}{2}\left.\frac{\partial^2}{\partial\eta^2} E'[\rho',\tilde{\rho}']\right|_{\eta=0}, \label{eq:m1EDF}
\end{align}
where the particle-hole and particle-particle densities $\rho'$ and $\tilde{\rho}'$ are constructed from the perturbed state $|\Phi'\rangle$.

\subsection{Isoscalar operator}

Equation (\ref{eq:perturbation}) can be regarded as a transformation of the wave function.
For an isoscalar operator $\hat{F}^{\rm IS}$, this is nothing but a local 
gauge transformation \cite{Blaizot-Ripka,PhysRevC.52.1827,PhysRevC.78.044326,PhysRevC.84.064303}. 
The local gauge transformation changes the particle-hole and particle-particle density matrices as \cite{PhysRevC.69.014316}
\begin{align}
  \hat{\rho}'(\rb s,\rbp s';t) &= e^{i\eta \alpha [f(\rb)- f(\rbp)]} \hat{\rho}(\rb s,\rbp s';t), \\
  \hat{\rhot}'(\rb s,\rbp s';t)&= e^{i\eta \alpha[f(\rb)+f(\rbp)]} \hat{\rhot}(\rb s,\rbp s';t),
\end{align}
and the nonlocal densities transform as
\begin{align}
  \rho'_t(\rb,\rbp) &= e^{i\eta\alpha[f(\rb) - f(\rbp)]} \rho_t(\rb,\rbp), \\
  \bm{s}'_t(\rb,\rbp) &= e^{i\eta\alpha[f(\rb) - f(\rbp)]} \bm{s}_t(\rb,\rbp), \\
  \rhot'_t(\rb,\rbp) &= e^{i\eta\alpha[f(\rb) + f(\rbp)]} \tilde{\rho}_t(\rb,\rbp), \\
  \tilde{\bm{s}}'_t(\rb,\rbp) &= e^{i\eta\alpha[f(\rb) + f(\rbp)]} \tilde{\bm{s}}_t(\rb,\rbp).
\end{align}
In analogy with the Galilean transformation, a local momentum field can be defined as
\begin{align}
  \bm{p}(\rb) = \eta \alpha \bm{\nabla} f(\rb).
\end{align}
The local densities in the EDF transform as
\begin{align}
  \rho_k' &= \rho_k, \label{eq:rho} \\
  \tau_k' &= \tau_k + 2 \bm{p}\cdot \bm{j}_k + \bm{p}^2 \rho_k,  \label{eq:tau} \\
  \bm{s}'_k &= \bm{s}_k, \\
  \bm{T}'_k &= \bm{T}_k + 2 \bm{p}\cdot {\sf J}_k  + \bm{p}^2 \bm{s}_k, \\
  \bm{j}_k' &= \bm{j}_k + \bm{p}\rho_k, \\
  \bm{F}'_k &= \bm{F}_k + \bm{p} J_k + {\sf J}_k \cdot \bm{p} + \bm{p}(\bm{p}\cdot \bm{s}_k), \\
     {\sf J}_k' &= {\sf J}_k + \bm{p} \otimes \bm{s}_k, \\
     \tilde{\rho}_t' &= e^{2i\eta\alpha f} \tilde{\rho}_t, \\
     \tilde{\tau}_t' &= e^{2i\eta\alpha f} (\tilde{\tau}_t + i\bm{p}\cdot\bm{\nabla}\tilde{\rho}_t - \bm{p}^2 \tilde{\rho}_t), \\
     \tilde{\sf J}_t' &= e^{2i\eta\alpha f}\tilde{\sf J}_t. \label{eq:Jpair}
\end{align}
For the local gauge invariant EDF, the transformation above does not change the EDF in Eq.~(\ref{eq:EDF}),
except for the kinetic-energy term. From Eq.~(\ref{eq:tau}), the kinetic-energy term transforms as
\begin{align}
  E_{\rm kin}' &=
  \frac{\hbar^2}{2m}\int d\rb \tau_0' \nonumber \\
  &=
  \frac{\hbar^2}{2m}\int d\rb \left[ \tau_0 + 2\eta\alpha(\bm{\nabla}f)\cdot\bm{j}_0 + \eta^2 \alpha^2(\nabla f)^2\rho_0\right].
\end{align}
From Eq.~(\ref{eq:m1EDF}), the term proportional to $\eta^2$ contributes to the energy-weighted sum rule of the isoscalar operator.
Then one has 
\begin{align}
  m_1(\hat{F}^{\rm IS}) = \frac{\hbar^2}{2m} \alpha^2 \int d\rb [ \nabla f(\rb)]^2 \rho_0(\rb)
\end{align}
for the local gauge invariant EDF.

This is the derivation of the Thouless theorem without using the Hamiltonian operator and double commutator.
Only the local gauge invariance property of the EDF is imposed in the derivation, and thus 
the existence or absence of the spin, spin-orbit, and density-dependent terms both in the particle-hole and pairing channels does not contribute to the energy-weighted sum rule as long as the EDF is local gauge invariant.
As for the pairing channel, local gauge invariant pairing EDF does not contribute to the energy-weighted sum rule.
Such local gauge invariant EDFs are not limited to the ones with the isoscalar density dependence considered in Eq.~(\ref{eq:pairforce}), but include isovector density dependence \cite{PhysRevC.80.064301} and the Fayans functional with particle-hole density-gradient dependence \cite{FAYANS199619}.

One can consider a general EDF that does not hold the local gauge invariance.
Without the local gauge invariance,
the transformation introduces additional terms; but when computing the energy of the transformed state, 
the densities of an even-even nucleus are used.
Therefore any time-odd densities included in the transformed EDF vanish.
The contribution from the spin-orbit and tensor terms produce terms proportional to the spin density $\bm{s}$,
and thus they do not contribute to the energy-weighted sum as well.
The Coulomb functionals are written with the proton local particle-hole densities only, and they are local gauge invariant.
Thus the possible contributions are from the $\rho_k\tau_k$ and $\bm{j}^2_k$ terms in the particle-hole EDF,
and ${\rm Re} \rhot_t^\ast\taut_t$ and ${\rm Re} \rhot_t^\ast\Delta\rhot_t$ terms in the pairing EDF.
The particle-hole part and pairing part of the EDF transform as
\begin{align}
  \int & d\rb 
  \chi_k[\rho_k',\tau_k',\cdots] \nonumber \\
  =& \int d\rb \left\{ \chi_k[\rho_k,\tau_k,\cdots] + (C^{\tau}_k + C^j_k) \bm{p}^2 \rho_k^2\right\}  ,\\
  \int & d\rb \tilde{\chi}_t[\tilde{\rho}'_t, \tilde{\rho}^{\prime\ast}_t, \tilde{\tau}'_t,\cdots, \rho_0'] \nonumber \\
  =& \int d\rb \left\{ \tilde{\chi}_t[\tilde{\rho}_t, \tilde{\rho}^{\ast}_t, \tilde{\tau}_t, \cdots, \rho_0]
  - (4 \tilde{C}_t^{\Delta\rho} + \tilde{C}_t^\tau )
   \bm{p}^2 |\tilde{\rho}_t|^2\right\},
\end{align}
where terms which are nonzero in time-reversal-symmetric even-even systems are kept.

The combinations of the coefficients $(C^{\tau}_k + C^j_k)$
and $(4\tilde{C}_t^{\Delta\rho} + \tilde{C}_t^\tau )$ 
show that these additional terms exist only when the local gauge symmetry of $\rho_k\tau_k-\bm{j}^2_k$  and/or
${\rm Re} (4\tilde{\rho}_t^\ast \Delta\rhot_t-\tilde{\rho}_t^\ast\tilde{\tau}_t)$
is broken.
By taking the terms that are second order in $\eta$ and performing the integration, the Thouless theorem for the isoscalar operator in the nuclear EDF is derived:
\begin{align}
  m_1(\hat{F}^{\rm IS}) &= \alpha^2 
  \int d\rb [ \nabla f(\rb)]^2  \biggl\{ \frac{\hbar^2}{2m} \rho_0(\rb) \nonumber \\
  &\quad + \sum_{k=0}^1 (C^{\tau}_k + C^j_k) \rho_k(\rb)^2 \nonumber \\
  &\quad - \sum_{t=n,p} (4 \tilde{C}_t^{\Delta\rho} + \tilde{C}_t^\tau) |\rhot_t(\rb)|^2 \biggr\}.
  \label{eq:EWSR-IS}
\end{align}
Note that in Ref.~\cite{Lipparini1989103} it is discussed that
the sum rule is obtained by the exact cancellation of the potential contribution to the effective mass ($\rho\tau$ term)
and the isoscalar current-current interaction in the RPA level
for the system with $N=Z$ and without spin-orbit interaction.
The present derivation based on the local gauge transformation gives a unified view,
that includes the contribution from the local gauge symmetry breakings of the isovector current terms
and pairing EDF, and it shows that the local gauge symmetry breaking in the spin-orbit and tensor functionals do not play any roles in the energy-weighted sum rule of the isoscalar coordinate operators.

\subsection{Isovector operator}

The energy-weighted sum rule of the isovector operator
for the nuclear EDF can be derived by generating the fluctuation using the
isovector operator given in Eq.~(\ref{eq:iv}). Consider a corresponding transformation
with the isovector operator
\begin{align}
  |\Phi'_{\rm IV}\rangle = \exp \left[
    i \eta \sum_{i=1}^A \alpha_{t_i} f(\hat{\rb}_i) \tau^1(t_i) \right]|\Psi_{\rm HFB}\rangle. \label{eq:ivtrans}
\end{align}
Because this is not a local gauge transformation,
even the local gauge invariant EDF is not invariant under this transformation.

The density matrices transform with Eq.~(\ref{eq:ivtrans}) as 
\begin{align}
  \hat{\rho}'(\rb s,\rbp s';t) &= e^{i (2t)\eta\alpha_t[f(\rb)
    -f(\rbp)]} \hat{\rho}(\rb s,\rbp s';t), \\
  \hat{\tilde{\rho}}'(\rb s,\rbp s';t) &= e^{i(2t)\eta\alpha_t[f(\rb)
      +f(\rbp)]} \hat{\tilde{\rho}}(\rb s,\rbp s';t).
\end{align}
Then nonlocal densities of neutrons and protons transform as
\begin{align}
  \rho'_t(\rb,\rbp) &= e^{i(2t)\eta\alpha_t [f(\rb) 
      - f(\rbp)]} \rho_t(\rb,\rbp), \label{eq:nlrhoiv1}\\
  \bm{s}'_t(\rb,\rbp) &= e^{i(2t)\eta\alpha_t[f(\rb)
      -f(\rbp)]} \bm{s}_t(\rb,\rbp), \label{eq:nlrhoiv2}\\
  \rhot'_t(\rb,\rbp) &= e^{i(2t)\eta\alpha_t [f(\rb) 
      + f(\rbp)]} \rhot_t(\rb,\rbp), \\
  \sbt'_t(\rb,\rbp) &= e^{i(2t)\eta\alpha_t[f(\rb)
      +f(\rbp)]} \sbt_t(\rb,\rbp).
\end{align}
Note that the indices in Eqs.~(\ref{eq:nlrhoiv1}) and (\ref{eq:nlrhoiv2}) are $t$.
One defines local momentum fields of the neutron and proton,
\begin{align}
  \bm{p}_t(\rb) =(2t)\eta\alpha_t\bm{\nabla} f(\rb).
\end{align}
The transformation in Eq.~(\ref{eq:ivtrans}) does not mix the neutron and proton phases. Therefore the isoscalar and isovector local densities transform as
\begin{align}
  \rho_k' &= \rho_k, \\
  \tau_0' &= \tau_0 + (\bm{p}_n + \bm{p}_p)\cdot \bm{j}_0 + \frac{1}{2}(\bm{p}_n^2 + \bm{p}_p^2) \rho_0 \nonumber \\
  &\quad + (\bm{p}_n - \bm{p}_p)\cdot \bm{j}_1 + \frac{1}{2}(\bm{p}_n^2 - \bm{p}_p^2) \rho_1, \\
  \tau_1' &= \tau_1 + (\bm{p}_n + \bm{p}_p)\cdot \bm{j}_1 + \frac{1}{2}(\bm{p}_n^2 + \bm{p}_p^2) \rho_1 \nonumber \\
  &\quad + (\bm{p}_n - \bm{p}_p)\cdot \bm{j}_0 + \frac{1}{2}(\bm{p}_n^2 - \bm{p}_p^2) \rho_0, \\
  \bm{s}_k' &= \bm{s}_k, \\
  \bm{T}_0' &= \bm{T}_0 + (\bm{p}_n + \bm{p}_p)\cdot {\sf J}_0 + \frac{1}{2}(\bm{p}_n^2 + \bm{p}_p^2) \bm{s}_0 \nonumber \\
  &\quad + (\bm{p}_n - \bm{p}_p)\cdot {\sf J}_1 + \frac{1}{2}(\bm{p}_n^2 - \bm{p}_p^2) \bm{s}_1, \\
  \bm{T}_1' &= \bm{T}_1 + (\bm{p}_n + \bm{p}_p)\cdot {\sf J}_1 + \frac{1}{2}(\bm{p}_n^2 + \bm{p}_p^2) \bm{s}_1 \nonumber \\
  &\quad + (\bm{p}_n - \bm{p}_p)\cdot {\sf J}_0 + \frac{1}{2}(\bm{p}_n^2 - \bm{p}_p^2) \bm{s}_0, \\
  \bm{j}_0' &= \bm{j}_0 + \frac{1}{2} (\bm{p}_n +  \bm{p}_p) \rho_0 + \frac{1}{2}(\bm{p}_n - \bm{p}_p)\rho_1, \\
  \bm{j}_1' &= \bm{j}_1 + \frac{1}{2} (\bm{p}_n +  \bm{p}_p) \rho_1 + \frac{1}{2}(\bm{p}_n - \bm{p}_p)\rho_0, \\
  \bm{F}_0' &= \bm{F}_0 + \frac{1}{2}(\bm{p}_n + \bm{p}_p) J_0 + \frac{1}{2}(\bm{p}_n - \bm{p}_p) J_1 \nonumber \\
  &\quad + \frac{1}{2} {\sf J}_0 \cdot (\bm{p}_n + \bm{p}_p) + \frac{1}{2} {\sf J}_1 \cdot (\bm{p}_n - \bm{p}_p) \nonumber \\
  &\quad + \frac{1}{2} [ (\bm{p}_n\cdot \bm{s}_0) \bm{p}_n + (\bm{p}_p\cdot \bm{s}_0)\bm{p}_p \nonumber \\
 & \quad + (\bm{p}_n\cdot \bm{s}_1) \bm{p}_n - (\bm{p}_p\cdot \bm{s}_1)\bm{p}_p],\\
  \bm{F}_1' &= \bm{F}_1 + \frac{1}{2}(\bm{p}_n + \bm{p}_p) J_1 + \frac{1}{2}(\bm{p}_n - \bm{p}_p) J_0\nonumber \\
  &\quad + \frac{1}{2} {\sf J}_1 \cdot (\bm{p}_n + \bm{p}_p) + \frac{1}{2} {\sf J}_0 \cdot (\bm{p}_n - \bm{p}_p) \nonumber \\
  &\quad + \frac{1}{2} [ (\bm{p}_n\cdot \bm{s}_1) \bm{p}_n + (\bm{p}_p\cdot \bm{s}_1)\bm{p}_p \nonumber \\
  &\quad + (\bm{p}_n\cdot \bm{s}_0) \bm{p}_n - (\bm{p}_p\cdot \bm{s}_0)\bm{p}_p],\\
       {\sf J}_0' &=  {\sf J}_0 + \frac{1}{2} (\bm{p}_n + \bm{p}_p)\otimes \bm{s}_0
       + \frac{1}{2}(\bm{p}_n - \bm{p}_p)\otimes \bm{s}_1, \\
       {\sf J}_1' &=  {\sf J}_1 + \frac{1}{2} (\bm{p}_n + \bm{p}_p)\otimes \bm{s}_1
       + \frac{1}{2}(\bm{p}_n - \bm{p}_p)\otimes \bm{s}_0, \\
       \tilde{\rho}_t' &= e^{2i(2t)\eta\alpha_tf} \tilde{\rho}_t, \\
       \tilde{\tau}_t' &= e^{2i(2t)\eta\alpha_tf}(\tilde{\tau}_t + i\bm{p}_t\cdot\bm{\nabla}\tilde{\rho}_t - \bm{p}_t^2 \tilde{\rho}_t), \\
       \tilde{\sf J}_t' &= e^{2i(2t)\eta\alpha_t f}\tilde{\sf J}_t.
\end{align}
Then consider an EDF transformed with Eq.~(\ref{eq:ivtrans}):
\begin{align}
  E'[ -i\eta F^{20}, i\eta F^{02}]
  &= \int d\rb \biggl\{ \frac{\hbar^2}{2m} \tau_0'
  + \sum_{k=0}^1 \chi_k[ \rho_k', \tau_k', \cdots] \nonumber \\ &\quad +
  \sum_{t=n,p} \tilde{\chi}_t[ \tilde{\rho}_t', \tilde{\rho}_t'^\ast, \tilde{\tau}_t', \cdots, \rho_0'] \biggr\} \nonumber\\
&\quad  +O(\eta^3).
\end{align}
The kinetic-energy term transforms as
\begin{align}
  \frac{\hbar^2}{2m} \tau_0' &=
  \frac{\hbar^2}{2m}\left[ \tau_0 + (\bm{p}_n + \bm{p}_p)\cdot \bm{j}_0 + \frac{1}{2}(\bm{p}_n^2+ \bm{p}_p^2) \rho_0  \right. \nonumber \\
    &\quad \left.  + (\bm{p}_n - \bm{p}_p)\cdot \bm{j}_1 + \frac{1}{2}(\bm{p}_n^2 - \bm{p}_p^2) \rho_1 \right]. \label{eq:ivkin}
\end{align}
Again in the particle-hole part, as the time-reversal symmetry cancels most of the terms, only the terms from $\rho_k\tau_k$ and $\bm{j}_k^2$ generate time-even
contribution to the transformed EDF:
\begin{widetext}
\begin{align}
  \int d\rb \sum_{k=0}^1 \chi_k[\rho_k',\tau_k',\cdots] 
  =&
  \int d\rb \sum_{k=0}^1 \biggl\{ \chi_k[\rho_k,\tau_k, \cdots] 
  + C_k^\tau\left[
    \left( \rho'_k\tau'_k - \bm{j}_k^{\prime 2} \right) - \left( \rho_k\tau_k - \bm{j}_k^2 \right)\right]\biggr\}\nonumber\\
  =&   \int d\rb \biggl\{ \sum_{k=0}^1 \chi_k[\rho_k,\tau_k, \cdots] 
   +\frac{1}{4}(C^{\tau}_0 - C^{\tau}_1)
    (\bm{p}_n -  \bm{p}_p)^2 (\rho_0^2 - \rho_1^2)\biggr\}. \label{eq:ivph}
\end{align}
\end{widetext}
The fluctuation of the pairing EDF does not contribute because the neutron and proton terms are independent in the pairing EDF.
By taking the terms proportional to $\eta^2$ from Eqs.~(\ref{eq:ivkin}) and (\ref{eq:ivph}),
\begin{align}
    m_1(\hat{F}^{\rm IV}) &=  \int d\rb [\nabla f(\rb)]^2 \biggl\{
    \frac{\hbar^2}{2m}\left[\alpha_n^2 \rho_n(\rb) + \alpha_p^2 \rho_p(\rb) \right] \nonumber \\
    &\quad + 
  (C_0^{\tau} - C_1^{\tau}) (\alpha_n + \alpha_p)^2 \rho_n(\rb)\rho_p(\rb)\biggr\} \nonumber \\
    &= m_1^{\rm kin}(\hat{F}^{\rm IV})\left[ 1 + \kappa(\hat{F}^{\rm IV})\right].
\end{align}
The first term is the kinetic-energy contribution, 
and the ratio to the second term defines the isovector enhancement factor $\kappa(\hat{F}^{\rm IV})$:
\begin{align}
  m_1^{\rm kin}(\hat{F}^{\rm IV}) &=   \frac{\hbar^2}{2m} \int d\rb [\nabla f(\rb)]^2 \left[\alpha_n^2 \rho_n(\rb) + \alpha_p^2 \rho_p(\rb) \right], \\
  \kappa(\hat{F}^{\rm IV}) &= \frac{2m}{\hbar^2} (C_0^{\tau} - C_1^{\tau})(\alpha_n+\alpha_p)^2 \nonumber \\
  &\quad \times \frac{\int d\rb [\nabla f(\rb)]^2 \rho_n(\rb)\rho_p(\rb)}
  {\int d\rb [\nabla f(\rb)]^2 \left[\alpha_n^2 \rho_n(\rb) + \alpha_p^2 \rho_p(\rb)\right]}.
\end{align}
$\alpha_n=\alpha_p=1$ produces Eqs.~(6.32) and (6.38) in Ref.~\cite{Lipparini1989103}:
\begin{align}
  m_1^{\rm kin}(\hat{F}^{\rm IV}) &=   \frac{\hbar^2}{2m} \int d\rb [\nabla f(\rb)]^2 \rho_0(\rb), \\
  \kappa &= \frac{8m}{\hbar^2} (C_0^{\tau} - C_1^{\tau})
  \frac{\int d\rb [\nabla f(\rb)]^2 \rho_n(\rb)\rho_p(\rb)}
  {\int d\rb [\nabla f(\rb)]^2 \rho_0(\rb)}.
\end{align}
$\alpha_n=Z/A$ and $\alpha_p=N/A$ are often used, especially for the dipole operator, to remove the contribution of the center-of-mass motion.
In the case of the isovector dipole operators $f(\rb)=f^{\rm IV}_{1K}(\rb) (K=0, 1)$,
one has a model-independent kinetic contribution (Thomas-Reiche-Kuhn sum rule \cite{Thomas1925,Ladenburg1923,Kuhn1925})
\begin{align}
  m_1^{\rm kin}(\hat{F}^{\rm IV}_{1K}) &= \frac{\hbar^2}{2m} \frac{3}{4\pi} \frac{NZ}{A}, \\
  \kappa^{\rm IV}_{1K} &= \frac{2m}{\hbar^2} \frac{A}{NZ} (C^{\tau}_0 - C^{\tau}_1)
  \int d\rb  \rho_n(\rb) \rho_p(\rb).
\end{align}
If the EDF does not hold the local gauge invariance,
again all the additional terms to the energy-weighted sum rule in the particle-hole channel
come from $\rho_k\tau_k$ and $\bm{j}_k^2$ terms,
\begin{widetext}
\begin{align}
  \int d\bm{r} \sum_{k=0}^1 \left( C^{\tau}_k \rho_k' \tau_k' +  C^j_k \bm{j}_k^{\prime 2} \right)
  &= \int d\bm{r}\biggl\{ \sum_{k=0}^1 C^{\tau}_k \rho_k \tau_k + (C^{\tau}_0 - C^{\tau}_1) (\bm{p}_n - \bm{p}_p)^2 \rho_n \rho_p \nonumber \\
  & \quad + (C^{\tau}_0 + C^j_0 + C^{\tau}_1 + C^j_1) (\bm{p}^2_n \rho_n^2 + \bm{p}^2_p \rho_p^2)
  + 2(C^{\tau}_0 + C^j_0 - C^{\tau}_1 - C^j_1) \bm{p}_n\cdot\bm{p}_p \rho_n \rho_p\biggr\}. \label{eq:ivrhotauminusj2}
\end{align}
The second term in the right-hand side of Eq.~(\ref{eq:ivrhotauminusj2}) is the contribution to the enhancement factor. The third and fourth terms vanish when the EDF is local gauge invariant for $\rho_k\tau_k$ and $\bm{j}_k^2$ terms ($C_k^j= -C_k^\tau$).

The pairing EDF transforms as
\begin{align}
  \tilde{\chi}_t[\rhot_t',\rhot_t^{\prime\ast}, \tau_t', \cdots, \rho_0']
  = \tilde{\chi}_t[\rhot_t,\rhot_t^\ast, \tau_t, \cdots, \rho_0]
  - (4 \Ct_t^{\Delta\rho} + \Ct_t^\tau) \pb^2_t |\rhot_t|^2,
\end{align}
and produces contributions from the local gauge symmetry breaking.
The energy-weighted sum rule of an isovector operator for the nuclear EDF is then given by
\begin{align}
      m_1  (\hat{F}^{\rm IV}) &=
      \int d\rb [\nabla f(\rb)]^2 \biggl\{
      \frac{\hbar^2}{2m} \left[\alpha_n^2 \rho_n(\rb) + \alpha_p^2 \rho_p(\rb) \right] 
      + 
  (C_0^{\tau} - C_1^{\tau}) (\alpha_n + \alpha_p)^2\rho_n(\rb)\rho_p(\rb) \nonumber \\
  &\quad +\sum_{k=0}^1(C^{\tau}_k + C^j_k) \left[\alpha_n \rho_n(\rb) +(-1)^{k+1} \alpha_p \rho_p(\rb)\right]^2
   - \sum_{t=n,p}(4 \Ct_t^{\Delta\rho} + \Ct_t^\tau) \alpha_t^2 |\rhot_t(\rb)|^2\biggr\}.
  \label{eq:EWSR-IV}
\end{align}
\end{widetext}

\section{Finite-amplitude method} \label{sec:fam}

To check the expressions for the energy-weighted sum rules for the nuclear EDF derived in the
previous section, QRPA calculations based on the linear-response theory have been performed.
In this section the procedure to calculate the energy-weighted sum rule from the linear response theory is summarized.

The finite-amplitude method (FAM) for computing the linear response is performed \cite{nakatsukasa:024318,PhysRevC.84.014314}.
The FAM allows one to perform a linear response within nuclear DFT
for a given external field $\hat{F}$ with a complex frequency $\omega$. 
By solving the linearized time-dependent Hartree-Fock-Bogoliubov equations, 
the strength function $S(\hat{F},\omega)$ can be numerically evaluated by an iterative method.
The strength function is written in terms of the QRPA energies and strengths as
\begin{align}
  S(\hat{F},\omega) = - \sum_{\lambda (\Omega_\lambda>0)}
  \left\{
  \frac{ |\langle\lambda|\hat{F}|0\rangle|^2}{\Omega_\lambda - \omega}
  +
  \frac{ |\langle 0 |\hat{F}|\lambda\rangle|^2}{\Omega_\lambda + \omega}
  \right\}.
\end{align}
A contour integration is performed in the complex-energy plane
to evaluate the energy-weighted sum rule numerically 
\begin{align}
  m_1(\hat{F}) = \frac{1}{2\pi i}\int_{A_1} \omega S(\hat{F},\omega) d\omega, \label{eq:A1}
\end{align}
where the integration path is taken to include all the positive-energy poles in the strength function.
The contour consists of a half counterclockwise arc $A_1$ from $\omega=-iR_{A_1}$ to $iR_{A_1}$ centered at the origin
and a line on the imaginary axis from $\omega=iR_{A_1}$ to $\omega=-iR_{A_1}$, which encircles all the poles in the range of $0 < \Omega_\lambda < R_{A1}$. For a Hermitian operator $\hat{F}$ the integration along the imaginary axis vanishes, and Eq.~(\ref{eq:A1}) is derived.
I refer the reader Ref.~\cite{PhysRevC.91.044323} for a more detailed discussion on the complex-energy FAM for the sum rules.

\section{Comparison of sum-rule values} \label{sec:sumrule}

In the numerical comparison, I use functionals based on UNEDF1-HFB \cite{0954-3899-42-3-034024},
which contains only the time-even coupling constants in the particle-hole channel.
Thus the UNEDF1-HFB functional does not correspond to a specific Hamiltonian operator and breaks the local gauge invariance.
For the comparison of sum-rule values, the following five UNEDF1-HFB EDFs with the different time-odd terms are considered
\begin{enumerate}
\item full time-odd terms derived by assuming the relation between the time-even and time-odd couplings;
\item only isoscalar and isovector current terms in the local gauge invariant form ($C^j_0=-C^\tau_0$ and $C^j_1=-C^\tau_1$);
\item only the isovector current term ($C^j_0=0, C^j_1=-C^\tau_1$);
\item only the isoscalar current term ($C^j_0=-C^\tau_0, C^j_1=0$);
\item no time-odd terms ($C^j_0=C^j_1=0$).
\end{enumerate}

For the time-odd terms of the UNEDF1-HFB functional, the following relations are assumed
\begin{align}
  C_0^s[\rho_0] &= -\frac{2}{3} C_0^\rho[\rho_0] - C_1^\rho[\rho_0], \\
  C_1^s[\rho_1] &= -\frac{1}{3} C_0^\rho[\rho_0], \\
  C_0^{\Delta s} &= \frac{1}{8}[ C_0^\tau + 3C_1^\tau - 4(C_0^{\Delta\rho} + C_1^{\Delta\rho})], \\
  C_1^{\Delta s} &= \frac{1}{24}[ 3(C_0^\tau - C_1^\tau) - 4 (C_0^{\Delta\rho} + C_1^{\Delta\rho})], \\
  C_k^j &= - C_k^\tau, \\
  C_k^{\nabla j} &= C_k^{\nabla J},
\end{align}
and $C_k^T = C_k^{\nabla s} = C_k^F =0$.

The energy-weighted sum rule of the monopole ($K=0$), dipole ($K=0$ and 1), quadrupole ($K=0$, 1, and 2), and
octupole ($K=0, 1, 2,$ and 3) operators of isoscalar and isovector type is computed.
Expressions for the energy-weighted sum rule of these operators in the cylindrical coordinate system are summarized in Appendix \ref{sec:multipole}.
$\alpha = Z/A$ is used for the isoscalar operators and $\alpha_n=Z/A$ and $\alpha_p=N/A$ are used for the isovector operators.

The calculations are performed with the HFBTHO code \cite{Stoitsov200543, Stoitsov20131592, PEREZ2017363} and its FAM extension
for the nonaxial finite $K$ modes \cite{PhysRevC.92.051302}.
This version of the code uses linearized densities explicitly, and thus parameter $\eta$ in the FAM is not necessary in the numerical calculation.
$N_{\rm sh}=20$ harmonic-oscillator shells are used as the single-particle model space,
and $N_{\rm GH}=40$, $N_{\rm GL}=40$, and $N_{\rm leg}=80$ points are used for the Gauss quadratures.
A \\\\\\60 MeV pairing window is employed.
In the FAM calculation the integration radius is set to $R_{A1}= 200$ MeV, and the half arc $A_1$ is discretized with $300$ points.

$^{208}$Pb is chosen as a representative case of the spherical state without pairing, and
$^{166}$Dy as a case with prolate deformation and pairing ($\beta=0.33$, $\Delta_n=0.64$ MeV, and $\Delta_p=0.58$ MeV).
Tables~\ref{table:208PbIS} and \ref{table:208PbIV} compare energy-weighted sum rule of $^{208}$Pb computed from the Thouless theorem using the HFB state [Eqs.~(\ref{eq:EWSR-IS}) and (\ref{eq:EWSR-IV})] with the one using the complex-energy FAM [Eq.~(\ref{eq:A1})] within six digits, 
while Tables~\ref{table:166DyIS} and \ref{table:166DyIV} are the same comparison but with $^{166}$Dy.
In $^{208}$Pb, the sum rule of different $K$ value for the same multipole $L$ gives the same value because of the spherical symmetry,
while in $^{166}$Dy, the sum-rule values depend on $K$ due to the ground-state deformation.
The agreement between the expressions from the Thouless theorem and the values from the complex-energy FAM is excellent.
In the comparison of the two calculations with full time-odd terms and with the isoscalar and isovector current terms only,
it is also numerically concluded that other time-odd terms involving the spin densities and breaking of the local gauge invariance of the spin-orbit functional do not play any role on the energy-weighted sum rule of the isoscalar and isovector multipole operators, as expected in the HFB expressions, Eqs. (\ref{eq:EWSR-IS}) and (\ref{eq:EWSR-IV}).
From the ratio of the sum-rule values computed with the FAM to the one computed from the Thouless theorem,
the maximum discrepancies of the isoscalar and isovector operators are about 0.7\% and 0.4\% in $^{208}$Pb
and 1.2\% and 0.5\% in $^{166}$Dy.
The cases with the largest discrepancy in $^{208}$Pb and $^{166}$Dy are the $K=1$ isoscalar dipole modes.
The isoscalar dipole operators shown in Eqs.~(\ref{eq:ISD0}) and (\ref{eq:ISD1}) easily couple with the spurious mode.
The standard prescription to subtract the spurious translational component, discussed in Ref.~\cite{VANGIAI19811}, was used, while
it is known that it still lacks small corrections coming from the higher-order terms \cite{PhysRevC.23.2329}. Although any one-body operator can be applied for the sum rule expressions and the complex-energy FAM calculation, the translational spurious component can affect the accuracy. The $L$-odd modes have to break the reflection symmetry in the FAM calculation, and this generally causes lower accuracy. Except for the isoscalar dipole mode, the maximum discrepancy between the FAM and the HFB expression of the energy-weighted sum rules is 0.5\% in the isoscalar and isovector $K=3$ octupole modes of $^{166}$Dy.

Because the current-density terms do not change the HFB state, the difference in the energy-weighted sum-rule value between the calculations with/without the current-density terms shows the actual contribution of the local gauge symmetry breaking.
The effect of the isoscalar (isovector) current density is much larger than the other in the sum rule of the isoscalar (isovector) multipole operator. This is because the contribution of the isoscalar (isovector) current-density term to the energy-weighted sum rule of the isoscalar operator is proportional to the isoscalar (isovector) particle-hole local density squared in Eq.~(\ref{eq:EWSR-IS}), and the isoscalar particle-hole local density is generally much larger than the isovector density. For the isovector multipole operator, as seen in Eq.~(\ref{eq:EWSR-IV}),
the contribution of the isovector current-density term is from an isoscalar-type density squared (in phase with $\alpha_n$ and $\alpha_p$ weight factors), while that of the isoscalar current-density term is from an isovector-type (out of phase) density squared.

\begin{table*}[h]
  \caption{
    Energy-weighted sum rule of the isoscalar monopole (ISM), dipole (ISD), quadrupole (ISQ), and octupole (ISO) operators computed from Eq.~(\ref{eq:EWSR-IS}) in the HFB states and the complex-energy FAM for $^{208}$Pb. UNEDF1-HFB functional is employed.
    Five choices for the time-odd coupling constants are listed.
    The units are in MeV fm$^x$ where $x=4,2,4$, and $6$ for $L=0,1,2$, and $3$ modes, respectively, and the scales are in parentheses.
    \label{table:208PbIS}      
}
  \begin{ruledtabular}
    \begin{tabular}{llcccccccccc}
      &&  \multicolumn{2}{c}{ISM($K=0$)} & \multicolumn{2}{c}{ISD($K=0$)} & \multicolumn{2}{c}{ISD($K=1$)} & \multicolumn{2}{c}{ISQ($K=0$)} & \multicolumn{2}{c}{ISQ($K=1$)}  \\
      && HFB & FAM & HFB & FAM & HFB & FAM & HFB & FAM &HFB & FAM \\
      && $(10^4)$ & $(10^4)$ & $(10^5)$ & $(10^5)$ & $(10^5)$ & $(10^5)$ & $(10^4)$ & $(10^4)$ & $(10^4)$ & $(10^4)$ \\ \hline
      full T-odd &                        & 8.29475 & 8.31797 & 2.90910 & 2.93022 & 2.90910 & 2.93023 & 1.65019 & 1.65219 & 1.65019 & 1.65222 \\ 
      $C_0^j=-C^\tau_0$ & $C_1^j=-C^\tau_1$ & 8.29475 & 8.31797 & 2.90910 & 2.93033 & 2.90910 & 2.93034 & 1.65019 & 1.65247 & 1.65019 & 1.65253 \\
      $C_0^j=0$ & $C_1^j=-C^\tau_1$        & 8.65237 & 8.67624 & 3.01255 & 3.03436 & 3.01255 & 3.03436 & 1.72133 & 1.72337 & 1.72133 & 1.72332 \\
      $C_0^j=-C^\tau_0$ & $C_1^j=0$        & 8.25788 & 8.28147 & 2.90909 & 2.91800 & 2.90909 & 2.91800 & 1.64285 & 1.64518 & 1.64285 & 1.64525 \\
      $C_0^j=0$ & $C_1^j=0$               & 8.61551 & 8.63972 & 3.01254 & 3.02201 & 3.01254 & 3.02201 & 1.71400 & 1.71607 & 1.71400 & 1.71607 \\ \\
      &&  \multicolumn{2}{c}{ISQ($K=2$)} & \multicolumn{2}{c}{ISO($K=0$)} & \multicolumn{2}{c}{ISO($K=1$)} & \multicolumn{2}{c}{ISO($K=2$)} & \multicolumn{2}{c}{ISO($K=3$)}  \\
      && HFB & FAM & HFB & FAM &HFB & FAM &HFB & FAM &HFB & FAM \\
      && $(10^4)$ & $(10^4)$ & $(10^6)$ & $(10^6)$ & $(10^6)$ & $(10^6)$ & $(10^6)$ & $(10^6)$ & $(10^6)$ & $(10^6)$ \\ \hline
      full T-odd &                      & 1.65019 & 1.65221 & 1.36753 & 1.37042 & 1.36753 & 1.37042 & 1.36753 & 1.37042 & 1.36753 & 1.37042 \\
      $C_0^j=-C^\tau_0$&$C_1^j=-C^\tau_1$ & 1.65019 & 1.65248 & 1.36753 & 1.37053 & 1.36753 & 1.37053 & 1.36753 & 1.37053 & 1.36753 & 1.37053 \\
       $C_0^j=0$&$C_1^j=-C^\tau_1$        & 1.72134 & 1.72336 & 1.41508 & 1.41820 & 1.41508 & 1.41820 & 1.41508 & 1.41820 & 1.41508 & 1.41820 \\
      $C_0^j=-C^\tau_0$&$C_1^j=0$         & 1.64285 & 1.64521 & 1.36204 & 1.36511 & 1.36204 & 1.36512 & 1.36204 & 1.36511 & 1.36204 & 1.36511 \\
       $C_0^j=0$&$C_1^j=0$               & 1.71400 & 1.71607 & 1.40959 & 1.41278 & 1.40959 & 1.41278 & 1.40959 & 1.41278 & 1.40959 & 1.41278
    \end{tabular}
  \end{ruledtabular}
\end{table*}
     
\begin{table*}[h]
  \caption{Energy-weighted sum rule of the isovector monopole (IVM), dipole (IVD), quadrupole (IVQ), and octupole (IVO) operators for $^{208}$Pb. The HFB values are evaluated using Eq.~(\ref{eq:EWSR-IV}).\label{table:208PbIV}}
  \begin{ruledtabular}
    \begin{tabular}{llcccccccccc}
      &&  \multicolumn{2}{c}{IVM($K=0$)} & \multicolumn{2}{c}{IVD($K=0$)} & \multicolumn{2}{c}{IVD($K=1$)} & \multicolumn{2}{c}{IVQ($K=0$)} & \multicolumn{2}{c}{IVQ($K=1$)}  \\
      && HFB & FAM & HFB & FAM &HFB & FAM &HFB & FAM &HFB & FAM \\
      && $(10^5)$ & $(10^5)$ & $(10^2)$ & $(10^2)$ & $(10^2)$ & $(10^2)$ & $(10^4)$ & $(10^4)$ & $(10^4)$ & $(10^4)$ \\ \hline
      full T-odd &                       & 1.46148 & 1.46576 & 2.93309 & 2.92660 & 2.93309 & 2.92640 & 2.90752 & 2.91122 & 2.90752 & 2.91117 \\ 
      $C_0^j=-C^\tau_0$&$C_1^j=-C^\tau_1$  & 1.46148 & 1.46576 & 2.93309 & 2.92541 & 2.93309 & 2.92530 & 2.90752 & 2.91123 & 2.90752 & 2.91114 \\
       $C_0^j=0$&$C_1^j=-C^\tau_1$        & 1.46161 & 1.46592 & 2.93328 & 2.92560 & 2.93328 & 2.92547 & 2.90778 & 2.91151 & 2.90778 & 2.91141 \\ 
      $C_0^j=-C^\tau_0$&$C_1^j=0$         & 1.31977 & 1.32456 & 2.60286 & 2.60392 & 2.60286 & 2.60394  & 2.62559 & 2.62968  & 2.62559 & 2.62968 \\
      $C_0^j=0$&$C_1^j=0$                 & 1.31989 & 1.32471 & 2.60305 & 2.60412 & 2.60305 & 2.60413 & 2.62584 & 2.62995 & 2.62584 &  2.62995\\ \\
      &&  \multicolumn{2}{c}{IVQ($K=2$)} & \multicolumn{2}{c}{IVO($K=0$)} & \multicolumn{2}{c}{IVO($K=1$)} & \multicolumn{2}{c}{IVO($K=2$)} & \multicolumn{2}{c}{IVO($K=3$)}  \\ 
      && HFB & FAM & HFB & FAM &HFB & FAM &HFB & FAM &HFB & FAM \\
      && $(10^4)$ & $(10^4)$ & $(10^6)$ & $(10^6)$ & $(10^6)$ & $(10^6)$ & $(10^6)$ & $(10^6)$ & $(10^6)$ & $(10^6)$ \\ \hline
      full T-odd &                       & 2.90752 & 2.91122 & 2.30407 & 2.31071 & 2.30407 & 2.31070 & 2.30407 & 2.31071 & 2.30407 & 2.31071 \\
      $C_0^j=-C^\tau_0$&$C_1^j=-C^\tau_1$  & 2.90752 & 2.91124 & 2.30407 & 2.31075 & 2.30407 & 2.31074 & 2.30407 & 2.31074 & 2.30407 & 2.31075 \\
       $C_0^j=0$&$C_1^j=-C^\tau_1$        & 2.90778 & 2.91151 & 2.30431 & 2.31104 & 2.30431 & 2.31103 & 2.30431 & 2.31103 & 2.30431 & 2.31104 \\
      $C_0^j=-C^\tau_0$&$C_1^j=0$         & 2.62559 & 2.62968 & 2.11636 & 2.12256 & 2.11636 & 2.12256 & 2.11636 & 2.12256 & 2.11636 & 2.12256 \\
       $C_0^j=0$&$C_1^j=0$               & 2.62585 & 2.62995 & 2.11660 & 2.12285 & 2.11660 & 2.12285 & 2.11661 & 2.12285 & 2.11660 & 2.12285
    \end{tabular}
  \end{ruledtabular}
\end{table*}

\begin{table*}[h]
  \caption{Energy-weighted sum rule of the isoscalar multipole operators for $^{166}$Dy.\label{table:166DyIS}}
  \begin{ruledtabular}
    \begin{tabular}{llcccccccccc}
      &&  \multicolumn{2}{c}{ISM($K=0$)} & \multicolumn{2}{c}{ISD($K=0$)} & \multicolumn{2}{c}{ISD($K=1$)} & \multicolumn{2}{c}{ISQ($K=0$)} & \multicolumn{2}{c}{ISQ($K=1$)}  \\
      && HFB & FAM & HFB & FAM &HFB & FAM &HFB & FAM &HFB & FAM \\
      && $(10^4)$ & $(10^4)$ & $(10^5)$ & $(10^5)$ & $(10^5)$ & $(10^5)$ & $(10^4)$ & $(10^4)$ & $(10^4)$ & $(10^4)$ \\ \hline
      full T-odd &                      & 6.09250 & 6.11890 & 3.07815 & 3.09868 & 1.52435 & 1.54264 & 1.46544 & 1.46837 & 1.33875 & 1.34136 \\
      $C_0^j=-C^\tau_0$&$C_1^j=-C^\tau_1$ & 6.09250 & 6.11891 & 3.07815 & 3.09873 & 1.52435 & 1.54273 & 1.46544 & 1.46839 & 1.33875 & 1.34139 \\
       $C_0^j=0$&$C_1^j=-C^\tau_1$        & 6.35181 & 6.37858 & 3.19099 & 3.21158 & 1.57586 & 1.59336 & 1.52926 & 1.53223 & 1.39645 & 1.39910 \\ 
      $C_0^j=-C^\tau_0$&$C_1^j=0$         & 6.06814 & 6.09444 & 3.07814 & 3.08693 & 1.52435 & 1.53668 & 1.45970 & 1.46265 & 1.33346 & 1.33610 \\
       $C_0^j=0$&$C_1^j=0$               & 6.32745 & 6.35411 & 3.19098 & 3.19960 & 1.57565 & 1.58738 & 1.52351 & 1.52648 & 1.39116 & 1.39380 \\ \\
      &&  \multicolumn{2}{c}{ISQ($K=2$)} & \multicolumn{2}{c}{ISO($K=0$)} & \multicolumn{2}{c}{ISO($K=1$)} & \multicolumn{2}{c}{ISO($K=2$)} & \multicolumn{2}{c}{ISO($K=3$)}  \\
      && HFB & FAM & HFB & FAM &HFB & FAM &HFB & FAM &HFB & FAM \\
      && $(10^3)$ & $(10^3)$ & $(10^6)$ & $(10^6)$ & $(10^6)$ & $(10^6)$ & $(10^5)$ & $(10^5)$ & $(10^5)$ & $(10^5)$ \\ \hline
      full T-odd &                      & 9.58688 & 9.60887 & 1.18672 & 1.19139 & 1.10894 & 1.11269 & 8.90831 & 8.94575 & 5.78137 & 5.81095 \\
      $C_0^j=-C^\tau_0$&$C_1^j=-C^\tau_1$ & 9.58688 & 9.60907 & 1.18672 & 1.19141 & 1.10894 & 1.11276 & 8.90831 & 8.94645 & 5.78137 & 5.81155 \\
       $C_0^j=0$&$C_1^j=-C^\tau_1$       & 9.98050 & 10.0028 & 1.22880 & 1.23350 & 1.14778 & 1.15163 & 9.20838 & 9.24757 & 5.96444 & 5.99449 \\
      $C_0^j=-C^\tau_0$&$C_1^j=0$        & 9.54736 & 9.56949 & 1.18218 & 1.18710 & 1.10477 & 1.10883 & 8.87621 & 8.91577 & 5.75952 & 5.79003 \\ 
       $C_0^j=0$&$C_1^j=0$               & 9.94097 & 9.96316 & 1.22426 & 1.22914 & 1.14361 & 1.14773 & 9.17628 & 9.21599 & 5.94258 & 5.97323
      \end{tabular}
  \end{ruledtabular}
\end{table*}

\begin{table*}[h]
  \caption{Energy-weighted sum rule of the isovector multipole operators for $^{166}$Dy. \label{table:166DyIV}}
  \begin{ruledtabular}
    \begin{tabular}{llcccccccccc}
      &&  \multicolumn{2}{c}{IVM($K=0$)} & \multicolumn{2}{c}{IVD($K=0$)} & \multicolumn{2}{c}{IVD($K=1$)} & \multicolumn{2}{c}{IVQ($K=0$)} & \multicolumn{2}{c}{IVQ($K=1$)}  \\
      && HFB & FAM & HFB & FAM &HFB & FAM &HFB & FAM &HFB & FAM \\
      && $(10^4)$ & $(10^4)$ & $(10^2)$ & $(10^2)$ & $(10^2)$ & $(10^2)$ & $(10^4)$ & $(10^4)$ & $(10^4)$ & $(10^4)$ \\ \hline
      full T-odd &                      & 10.5723 & 10.6141 & 2.34288 & 2.34783 & 2.34288 & 2.34584 & 2.55819 & 2.56284 & 2.33075 & 2.33496 \\
      $C_0^j=-C^\tau_0$&$C_1^j=-C^\tau_1$ & 10.5723 & 10.6141 & 2.34288 & 2.34797 & 2.34288 & 2.34599 & 2.55819 & 2.56284 & 2.33075 & 2.33496 \\
      $C_0^j=0$&$C_1^j=-C^\tau_1$        & 10.5734 & 10.6154 & 2.34306 & 2.34815 & 2.34306 & 2.34618 & 2.55845 & 2.56312 & 2.33098 & 2.33520 \\
      $C_0^j=-C^\tau_0$&$C_1^j=0$         & 9.55521 & 9.59796 & 2.08108 & 2.08912 & 2.08108 & 2.08758 & 2.30750 & 2.31253 & 2.10422 & 2.10859 \\
      $C_0^j=0$&$C_1^j=0$               & 9.55624 & 9.59918 & 2.08126 & 2.08930 & 2.08126 & 2.08777 & 2.30775 & 2.31280 & 2.10445 & 2.10884 \\ \\
      &&  \multicolumn{2}{c}{IVQ($K=2$)} & \multicolumn{2}{c}{IVO($K=0$)} & \multicolumn{2}{c}{IVO($K=1$)} & \multicolumn{2}{c}{IVO($K=2$)} & \multicolumn{2}{c}{IVO($K=3$)}  \\
      && HFB & FAM & HFB & FAM &HFB & FAM &HFB & FAM &HFB & FAM \\
      && $(10^4)$ & $(10^4)$ & $(10^6)$ & $(10^6)$ & $(10^6)$ & $(10^6)$ & $(10^6)$ & $(10^6)$ & $(10^5)$ & $(10^5)$ \\ \hline
      full T-odd &                      & 1.64840 & 1.65188 & 2.15229 & 2.16107 & 1.96199 & 1.96768 & 1.47491 & 1.48066 & 9.42464 & 9.46666 \\
      $C_0^j=-C^\tau_0$&$C_1^j=-C^\tau_1$ & 1.64840 & 1.65188 & 2.15229 & 2.16103 & 1.96199 & 1.96753 & 1.47491 & 1.48065 & 9.42464 & 9.46689 \\
       $C_0^j=0$&$C_1^j=-C^\tau_1$       & 1.64856 & 1.65205 & 2.15254 & 2.16129 & 1.96222 & 1.96778 & 1.47507 & 1.48039 & 9.42578 & 9.46813 \\
      $C_0^j=-C^\tau_0$&$C_1^j=0$        & 1.49440 & 1.49802 & 1.97080 & 1.97992 & 1.79885 & 1.80654 & 1.35767 & 1.36397 & 8.71254 & 8.75830 \\
       $C_0^j=0$&$C_1^j=0$              & 1.49455 & 1.49820 & 1.97105 & 1.98018 & 1.79908 & 1.80570 & 1.35784 & 1.36414 & 8.71368 & 8.75954
    \end{tabular}
  \end{ruledtabular}
  \end{table*}

\section{Conclusions \label{sec:conclusion}}

The expressions for the energy-weighted sum rule of the isoscalar and isovector coordinate operators 
are derived for the case of the nuclear DFT where the EDF does not correspond to a Hamiltonian.

The importance of the local gauge invariance of the nuclear EDF for evaluating the energy-weighted sum rule of these operators is discussed.
For time-reversal symmetric even-even systems, the local gauge invariance of the $\rho_k\tau_k - \bm{j}_k^2$ term
in the particle-hole channel and  ${\rm Re}(4\rhot^\ast_t\Delta\rhot_t - \rhot^\ast_t\taut_t)$ in the pairing channel
is responsible for the energy-weighted sum-rule value of the conventional Thouless theorem, while the local gauge invariance of the other terms such as spin-orbit and tensor does not play any role in the energy-weighted sum rule of the multipole operators.
The expressions for the energy-weighted sum-rule values are compared with the QRPA calculations with the complex-energy FAM, and expressions derived are both analytically and numerically justified.

The ratio of energy-weighted and inverse-energy-weighted sum rule is useful for estimating the giant resonance energy. The present derivation establishes  the efficient evaluation of the sum-rule ratio for the nuclear EDF that does not correspond to a Hamiltonian, as the dielectric theorem is available for the nuclear EDF to evaluate the inverse-energy-weighted sum rule \cite{PhysRevC.79.054329}.

The local gauge invariance of $\rho_k\tau_k-\bm{j}_k^2$ is related to the Galilean invariance, and thus almost all the practical nuclear EDFs should hold it. However, the present derivation of the Thouless theorem is also applicable to
other kinds of operators such as spin and isospin.
The energy-weighted sum rule of the spin operators is related to the spin-orbit and tensor energy terms
\cite{PhysRev.130.1525,ZAMICK198187}. It will be very useful to derive the expression for the energy-weighted sum rule of the spin and spin-multipole operators for better understanding of the spin-orbit and tensor terms in nuclear EDFs.

Extensions to non-Hermitian operators such as charge-exchange and pair transfer excitation, and the derivation of the cubic energy-weighted sum rule within the nuclear DFT, are other challenging future subjects.

\section*{Acknowledgments}

Discussions with Markus Kortelainen and Witold Nazarewicz are acknowledged.
This work is supported by the JSPS KAKENHI Grants No. 16K17680 and No. 17H05194, and
the JSPS-NSFC Bilateral Program for the Joint Research Project on ``Nuclear mass and life for unravelling mysteries of r-process.''
Numerical calculations were performed at the COMA (PACS-IX) 
and Oakforest-PACS Systems through the Multidisciplinary Cooperative Research Program of the Center for Computational Sciences, University of Tsukuba.

\appendix

\section{Densities} \label{sec:densities}

The particle-hole and particle-particle density matrices are given by
\begin{align}
  \hat{\rho}(\rb s,\rbp s';t) &= \langle \hat{c}^\dag_{\rbp s't} \hat{c}_{\rb st}\rangle,\\
  \hat{\tilde{\rho}}(\rb s,\rbp s';t) &= -2s'\langle \hat{c}_{\rbp-s't} \hat{c}_{\rb st}\rangle,
\end{align}
where $\hat{c}^\dag$ and $\hat{c}$ are nucleon creation and annihilation operators.
The nonlocal densities are expressed in terms of the density matrices as
\begin{align}
  \rho_k(\rb,\rbp) &= \sum_{st}\hat{\rho}(\rb s,\rbp s;t) \tau^k(t), \\
  \bm{s}_k(\rb,\rbp) &= \sum_{ss't}\hat{\rho}(\rb s,\rbp s';t) \bm{\sigma}_{s's}\tau^k(t), \\
  \rhot_t(\rb,\rbp) &= \sum_{s}\hat{\tilde{\rho}}(\rb s,\rbp s;t), \\
  \sbt_t(\rb,\rbp) &= \sum_{ss'}\hat{\tilde{\rho}}(\rb s,\rbp s';t) \bm{\sigma}_{s's},
\end{align}
where $\tau^k(t) = 1$ for $k=0$, and $2t$ for $k=1$.

All the local densities appear in the nuclear EDF are derived from nonlocal densities as
\begin{align}
  \rho_k(\rb) &= \rho_k(\rb,\rb), \\
  \tau_k(\rb) &= [(\bm{\nabla}\cdot \bm{\nabla}') \rho_k(\rb,\rbp)]_{\rb=\rbp}, \\
  {\sf J}_k(\rb) &= \frac{1}{2i}[ (\bm{\nabla} - \bm{\nabla}')\otimes \bm{s}_k(\rb,\rbp)]_{\rb=\rbp},\\
  \bm{s}_k(\rb) &= \bm{s}_k(\rb,\rb), \\
  \bm{T}_k(\rb) &= [(\bm{\nabla}\cdot \bm{\nabla}') \bm{s}_k(\rb,\rbp)]_{\rb=\rbp}, \\
  \bm{j}_k(\rb) &= \frac{1}{2i}[ (\bm{\nabla} - \bm{\nabla}') \rho_k(\rb,\rbp)]_{\rb=\rbp}, \\
  \bm{F}_k(\rb) &= \frac{1}{2}[ ( \bm{\nabla} \otimes \bm{\nabla}' + \bm{\nabla}'\otimes \bm{\nabla})\cdot \bm{s}_k(\rb,\rbp)]_{\rb=\rbp}, \\
  \rhot_t(\rb) &= \rhot_t(\rb,\rb), \\  
  \taut_t(\rb) &= [(\bm{\nabla}\cdot \bm{\nabla}') \rhot_t(\rb,\rbp)]_{\rb=\rbp}, \\
  \Jsft_t(\rb) &= \frac{1}{2i}[ (\bm{\nabla} - \bm{\nabla}')\otimes \sbt_t(\rb,\rbp)]_{\rb=\rbp},
\end{align}
and tensor densities can be decomposed into
\begin{align}
  J_k(\rb) &= \sum_{a} {\sf J}_{kaa}(\rb), \\
  \bm{J}_{ka}(\rb) &= \sum_{bc} \varepsilon_{abc}{\sf J}_{kbc}(\rb), \\
  \underline{\sf J}_{kab}(\rb) &= \frac{1}{2}{\sf J}_{kab}(\rb) + \frac{1}{2}{\sf J}_{kba}(\rb) - \frac{1}{3}J_k(\rb)\delta_{ab}, \\
    \Jt_k(\rb) &= \sum_{a} \Jsft_{kaa}(\rb), \\
  \Jbt_{ka}(\rb) &= \sum_{bc} \varepsilon_{abc}\Jsft_{kbc}(\rb), \\
  \underline{\Jsft}_{kab}(\rb) &= \frac{1}{2}\Jsft_{kab}(\rb) + \frac{1}{2}\Jsft_{kba}(\rb) - \frac{1}{3}\Jt_k(\rb)\delta_{ab}.
\end{align}

\section{Energy-weighted sum rule expressions for multipole operators} \label{sec:multipole}

The expressions for the energy-weighted sum rule of the multipole operators up to $L=3$ in cylindrical coordinates are summarized in this section.
The multipole operators $f_{LK}(\rb)$ are expressed using $x=\rho\cos\phi, y=\rho\sin\phi$ as
\begin{align}  
  f_{00}(\rb) &= r^2 = \rho^2 + z^2, \\
  f_{10}^{\rm IS}(\rb) &= r^3 Y_{10} - \eta_{10} rY_{10} = \sqrt{\frac{3}{4\pi}} ( z^3 + \rho^2 z - \eta_{10} z), \label{eq:ISD0} \\
  f_{10}^{\rm IV}(\rb) &= r Y_{10} = \sqrt{\frac{3}{4\pi}} z, \\
  f_{11}^{\rm IS}(\rb) &= (r^3 - \eta_{11}r)\frac{ Y_{11}- Y_{1-1}}{\sqrt{2}}\nonumber \\
  &= -\sqrt{\frac{3}{16\pi}}\rho( \rho^2+z^2 - \eta_{11})( e^{i\phi} + e^{-i\phi}), \label{eq:ISD1} \\
  f_{11}^{\rm IV}(\rb) &= \frac{r(Y_{11} - Y_{1-1})}{\sqrt{2}} = -\sqrt{\frac{3}{16\pi}}\rho
  ( e^{i\phi} + e^{-i\phi}), \\
  f_{20}(\rb) &= r^2 Y_{20}
  = \sqrt{\frac{5}{16\pi}} (2z^2 - \rho^2),\\
  f_{21}(\rb) &= -\frac{r^2 (Y_{21} - Y_{2-1})}{\sqrt{2}} =
  \sqrt{\frac{15}{16\pi}}\rho z(e^{i\phi}+e^{-i\phi}), \\
  f_{22}(\rb) &= \frac{r^2 (Y_{22} + Y_{2-2})}{\sqrt{2}} = \sqrt{\frac{15}{64\pi}} \rho^2
  (e^{2i\phi} + e^{-2i\phi}),\\
  f_{30}^{\rm IS}(\rb) &= r^3 Y_{30} - \eta_{30}r Y_{10} \nonumber \\
  &= \sqrt{ \frac{7}{16\pi}} ( 2z^3 - 3\rho^2 z - \eta_{30}'z), \\
  f_{30}^{\rm IV}(\rb) &= r^3 Y_{30} = \sqrt{\frac{7}{16\pi}}z(2z^2 - 3 \rho^2), \\
  f_{31}^{\rm IS}(\rb) &= \frac{r^3 (Y_{31}-Y_{3-1})}{\sqrt{2}} - \eta_{31} \frac{r (Y_{11}-Y_{1-1})}{\sqrt{2}} \nonumber \\
  &= \sqrt{\frac{21}{128\pi}} [ -4\rho z^2 + \rho^3 + \eta_{31}'\rho] (e^{i\phi} + e^{-i\phi}), \\
  f_{31}^{\rm IV}(\rb) &= 
  \frac{r^3 (Y_{31} - Y_{3-1})}{\sqrt{2}} \nonumber \\
  &=
  -\sqrt{\frac{21}{128\pi}} (4z^2 - \rho^2)\rho (e^{i\phi} + e^{-i\phi}),\\
  f_{32}(\rb) &= \frac{r^3 (Y_{32} + Y_{3-2})}{\sqrt{2}} = 
    \sqrt{\frac{105}{64\pi}}\rho^2 z(e^{2i\phi} + e^{-2i\phi}),\\
    f_{33}(\rb) &= \frac{r^3(Y_{33} - Y_{3-3})}{\sqrt{2}} = 
    -\sqrt{\frac{35}{128\pi}}\rho^3(e^{3i\phi} + e^{-3i\phi}).
\end{align}
The parameters $\eta_{LK}$ in the isoscalar dipole and octupole operators are given by
\cite{PhysRevC.23.2329,VANGIAI19811,yoshida:064316,PhysRevC.80.044324}
\begin{align}
  \eta_{10} &= \frac{1}{A} \int d\rb (3z^2 + \rho^2) \rho_0(\rb), \\
  \eta_{11} &= \frac{1}{A} \int d\rb (z^2 + 2\rho^2) \rho_0(\rb), \\
  \eta_{30}' &= \sqrt{\frac{12}{7}}\eta_{30} = \frac{1}{A} \int d\rb (6z^2 - 3\rho^2) \rho_0(\rb), \\
  \eta_{31}' &= \sqrt{\frac{8}{7}}\eta_{31} = \frac{1}{A} \int d\rb (4z^2 - 2\rho^2) \rho_0(\rb).
\end{align}  
The sum rules are written using the root-mean-square radius and deformation parameters
\begin{align}
   \langle r^2_{t}\rangle &= \frac{ \int d\rb(\rho^2+z^2) \rho_t(\rb) }
           { \int d\rb \rho_t(\rb)} = \frac{1}{N_t}\int d\rb(\rho^2+z^2) \rho_t(\rb), \\
           \langle r^2_{\rm tot}\rangle &= \frac{  \int d\rb(\rho^2+z^2) \rho_0(\rb) }
                   { \int d\rb \rho_0(\rb)} = \frac{ N\langle r^2_n\rangle + Z\langle r^2_p\rangle}{A},
\end{align}\begin{align}
  \beta_{2t} &= \sqrt{\frac{\pi}{5}} \frac{  \int d\rb(2z^2 - \rho^2) \rho_t(\rb) }
       { \int d\rb (\rho^2+z^2)\rho_t(\rb)} \nonumber \\
       &= \frac{1}{ N_t \langle r^2_t\rangle} \sqrt{\frac{\pi}{5}} \int d\rb(2z^2 - \rho^2) \rho_t(\rb),\\
       \beta_2 &=
\sqrt{\frac{\pi}{5}}
\frac{ \int d\bm{r} (2 z^2 - \rho^2) \rho_0(\bm{r})}
{ \int d\bm{r} (\rho^2+z^2) \rho_0(\bm{r})}
\nonumber \\
&=
\frac{1}{A\langle r^2_{\rm tot}\rangle}
\sqrt{\frac{\pi}{5}}
\int d\bm{r} (2 z^2 - \rho^2) \rho_0(\bm{r}). 
\end{align}
The energy-weighted sum rules of isoscalar multipole operators are written as
\begin{align}
  m_1(\hat{F}^{\rm IS}) = m_1^{\rm kin}(\hat{F}^{\rm IS})+  m_1^{\rm LGSB}(\hat{F}^{\rm IS}),
\end{align}
where the first term is from the kinetic-energy term, and the second term is from the local gauge symmetry breaking of the particle-hole and pairing EDF.
The expressions for the multipole operators are
  \begin{align}    
    m_1^{\rm kin}(\hat{F}^{\rm IS}_{00}) &= 4\left(\frac{Z}{A}\right)^2 \frac{\hbar^2}{2m} A\langle r^2_{\rm tot}\rangle, \\
    m_1^{\rm LGSB}(\hat{F}^{\rm IS}_{00}) &= 4\left(\frac{Z}{A}\right)^2
    \int d\rb (\rho^2+z^2)
    G_{\rm IS}^{\rm LGSB}(\bm{r}),
\end{align}\begin{align}
    m_1^{\rm kin}(\hat{F}^{\rm IS}_{10}) &= \frac{3}{4\pi}\left(\frac{Z}{A}\right)^2 \frac{\hbar^2}{2m} \nonumber \\
    \times &\left[
    \int d\rb (\rho^4 + 10\rho^2z^2 + 9z^4)\rho_0(\rb) - \eta_{10}^2 A \right], \\
    m_1^{\rm LGSB}(\hat{F}^{\rm IS}_{10}) &=   \frac{3}{4\pi}\left(\frac{Z}{A}\right)^2\int d\rb [\rho^4+10\rho^2z^2+9z^4
      \nonumber \\
      & \quad +\eta_{10}^2 - 2\eta_{10}(\rho^2+3z^2) ]
    G_{\rm IS}^{\rm LGSB}(\bm{r}), \\
    m_1^{\rm kin}(\hat{F}^{\rm IS}_{11}) &=
    \frac{3}{4\pi} \left(\frac{Z}{A}\right)^2
    \frac{\hbar^2}{2m} \nonumber \\
    \times & \left[ \int d\rb(5\rho^4 + 6\rho^2 z^2 + z^4)\rho_0(\rb) - \eta_{11}^2A \right], \\
    m_1^{\rm LGSB}(\hat{F}^{\rm IS}_{11}) &=
    \frac{3}{4\pi} \left(\frac{Z}{A}\right)^2
    \int d\rb [5\rho^4+6\rho^2z^2+z^4 \nonumber \\
      &\quad + \eta_{11}^2 - 2\eta_{11}(2\rho^2+z^2) ]
    G_{\rm IS}^{\rm LGSB}(\bm{r}),
\end{align}\begin{align}
    m_1^{\rm kin}(\hat{F}^{\rm IS}_{20}) &= \frac{5}{2\pi}\left(\frac{Z}{A}\right)^2\frac{\hbar^2}{2m} A\langle r^2_{\rm tot}\rangle
    \left( 1 + \sqrt{\frac{5}{4\pi}}\beta_2 \right), \\
    m_1^{\rm LGSB}(\hat{F}^{\rm IS}_{20}) &= \frac{5}{4\pi}\left(\frac{Z}{A}\right)^2 \int d\rb (\rho^2+4z^2)
    G_{\rm IS}^{\rm LGSB}(\bm{r}), \\
    m_1^{\rm kin}(\hat{F}^{\rm IS}_{21}) &= \frac{5}{2\pi}\left(\frac{Z}{A}\right)^2 \frac{\hbar^2}{2m} A\langle r^2_{\rm tot}\rangle \left( 1 + \sqrt{\frac{5}{16\pi}}\beta_{2}\right), \\
    m_1^{\rm LGSB}(\hat{F}^{\rm IS}_{21}) &= \frac{15}{8\pi}\left(\frac{Z}{A}\right)^2 \int d\rb (\rho^2+2z^2)
    G_{\rm IS}^{\rm LGSB}(\bm{r}), \\
    m_1^{\rm kin}(\hat{F}^{\rm IS}_{22}) &= \frac{5}{2\pi}\left(\frac{Z}{A}\right)^2 \frac{\hbar^2}{2m} A\langle r^2_{\rm tot}\rangle
    \left( 1 - \sqrt{\frac{5}{4\pi}}\beta_2 \right), \\
    m_1^{\rm LGSB}(\hat{F}^{\rm IS}_{22}) &= \frac{15}{4\pi}\left(\frac{Z}{A}\right)^2\int d\rb \rho^2
    G_{\rm IS}^{\rm LGSB}(\bm{r}),
  \end{align}\begin{align}
    m_1^{\rm kin}(\hat{F}^{\rm IS}_{30}) &=
    \frac{7}{16\pi} \left(\frac{Z}{A}\right)^2 \frac{\hbar^2}{2m} \nonumber \\
    &\quad \times\left[ 9\int d\rb
      (\rho^4 +  4z^4)\rho_0(\rb) - \eta_{30}^{\prime 2} A \right], \\
    m_1^{\rm LGSB}(\hat{F}^{\rm IS}_{30}) &=   \frac{7}{16\pi} \left(\frac{Z}{A}\right)^2\int d\rb[9\rho^4 + 36z^4 \nonumber\\
     &\quad + \eta_{30}^{\prime 2}- 2\eta_{30}'(- 3\rho^2 +6z^2)]G_{\rm IS}^{\rm LGSB}(\bm{r}), \\
    m_1^{\rm kin}(\hat{F}^{\rm IS}_{31}) &= 
    \frac{21}{32\pi}  \left(\frac{Z}{A}\right)^2 \frac{\hbar^2}{2m}\nonumber\\
    \times \biggl[ \int & d\rb  (5\rho^4 +  16\rho^2 z^2+ 16z^4)\rho_0(\rb)  -\eta_{31}^{\prime 2}A
      \biggr], \\
    m_1^{\rm LGSB}(\hat{F}^{\rm IS}_{31}) &=  \frac{21}{32\pi}  \left(\frac{Z}{A}\right)^2 \int d\rb [ 5\rho^4 + 16\rho^2 z^2+ 16z^4   \nonumber \\
      & \quad + \eta_{31}^{\prime 2} - 2\eta'_{31}( - 2\rho^2 + 4z^2)]
    G_{\rm IS}^{\rm LGSB}(\bm{r}), \\
    m_1^{\rm kin}(\hat{F}^{\rm IS}_{32}) &= \frac{105}{32\pi}\left(\frac{Z}{A}\right)^2 \frac{\hbar^2}{2m}
    \int d\rb( 8\rho^2 z^2+ \rho^4) \rho_0(\rb), \\
    m_1^{\rm LGSB}(\hat{F}^{\rm IS}_{32}) &= \frac{105}{32\pi}\left(\frac{Z}{A}\right)^2
    \int d\rb ( 8\rho^2 z^2 + \rho^4) 
    G_{\rm IS}^{\rm LGSB}(\bm{r}), \\    
    m_1^{\rm kin}(\hat{F}^{\rm IS}_{33}) &=  \frac{315}{32\pi} \left(\frac{Z}{A}\right)^2 \frac{\hbar^2}{2m}
    \int d\rb \rho^4 \rho_0(\rb), \\
    m_1^{\rm LGSB}(\hat{F}^{\rm IS}_{33}) &=  \frac{315}{32\pi} \left(\frac{Z}{A}\right)^2 \int d\rb \rho^4
    G_{\rm IS}^{\rm LGSB}(\bm{r}),
  \end{align}
  where
  \begin{align}
    G_{\rm IS}^{\rm LGSB}(\bm{r}) \equiv& \sum_{k=0}^1  (C_k^\tau + C_k^j) \rho_k^2(\rb) \nonumber \\
    & - \sum_{t=n,p} (4 \Ct_t^{\Delta\rho} + \Ct_t^\tau) |\rhot_t(\rb)|^2.
  \end{align}
  The isovector sum rules are expressed as the sum of the kinetic term, enhancement factor, and
  the contribution from the local gauge symmetry breaking of the EDF
\begin{align}
  m_1(\hat{F}^{\rm IV}) = m_1^{\rm kin}(\hat{F}^{\rm IV}) \left[ 1 + \kappa(\hat{F}^{\rm IV})\right] + m_1^{\rm LGSB}(\hat{F}^{\rm IV}).
  \end{align}
The terms for the multipole operators are given by
  \begin{align}
  m_1^{\rm kin}(\hat{F}^{\rm IV}_{00}) &= 4 \frac{\hbar^2}{2m} \frac{NZ}{A^2}
  \left( Z \langle r_n^2\rangle  + N \langle r_p^2 \rangle \right), \\
  m_1^{\rm kin}\kappa (\hat{F}^{\rm IV}_{00}) &= 4(C_0^\tau - C_1^\tau) \int d\rb (\rho^2 + z^2) \rho_n(\rb) \rho_p(\rb), \\
  m_1^{\rm LGSB}(\hat{F}^{\rm IV}_{00}) &= 4 \frac{1}{A^2}
\int d\rb (\rho^2 + z^2) G_{\rm IV}^{\rm LGSB}(\bm{r}),
 \end{align}\begin{align}
  m_1^{\rm kin}(\hat{F}^{\rm IV}_{10}) &= \frac{3}{4\pi} \frac{\hbar^2}{2m} \frac{NZ}{A}, \\
  m_1^{\rm kin}\kappa (\hat{F}^{\rm IV}_{10}) &= \frac{3}{4\pi}(C_0^\tau - C_1^\tau) \int d\rb\rho_n(\rb) \rho_p(\rb), \\
  m_1^{\rm LGSB}(\hat{F}^{\rm IV}_{10}) &= \frac{3}{4\pi} \frac{1}{A^2}
  \int d\rb G_{\rm IV}^{\rm LGSB}(\bm{r}), \\
  m_1^{\rm kin}(\hat{F}^{\rm IV}_{11}) &= \frac{3}{4\pi} \frac{\hbar^2}{2m} \frac{NZ}{A}, \\
  m_1^{\rm kin}\kappa (\hat{F}^{\rm IV}_{11}) &= \frac{3}{4\pi}(C_0^\tau - C_1^\tau) \int d\rb\rho_n(\rb) \rho_p(\rb), \\
  m_1^{\rm LGSB}(\hat{F}^{\rm IV}_{11}) &= \frac{3}{4\pi} \frac{1}{A^2}
  \int d\rb G_{\rm IV}^{\rm LGSB}(\bm{r}),
    \end{align}  \begin{align}
   m_1^{\rm kin}(\hat{F}^{\rm IV}_{20}) &= \frac{5}{2\pi} \frac{\hbar^2}{2m} \frac{NZ}{A^2}
   \Biggl[ Z \langle r_n^2\rangle \left( 1 + \sqrt{\frac{5}{4\pi}} \beta_{2n} \right) \nonumber \\
     &\quad
     + N \langle r_p^2 \rangle \left( 1 + \sqrt{\frac{5}{4\pi}} \beta_{2p}\right) \Biggr],\\
  m_1^{\rm kin}\kappa (\hat{F}^{\rm IV}_{20}) &= \frac{5}{4\pi}(C_0^\tau - C_1^\tau) \int d\rb (\rho^2 + 4z^2) \rho_n(\rb) \rho_p(\rb), \\
  m_1^{\rm LGSB}(\hat{F}^{\rm IV}_{20}) &= \frac{5}{4\pi} \frac{1}{A^2}
  \int d\rb (\rho^2 + 4z^2) G_{\rm IV}^{\rm LGSB}(\bm{r}), \\
  m_1^{\rm kin}(\hat{F}^{\rm IV}_{21}) &= \frac{5}{2\pi} \frac{\hbar^2}{2m} \frac{NZ}{A^2}
  \Biggl[
  Z\langle r_n^2\rangle \left( 1 + \sqrt{\frac{5}{16\pi}} \beta_{2n} \right) \nonumber \\
  &\quad +N\langle r_p^2\rangle \left( 1 + \sqrt{\frac{5}{16\pi}} \beta_{2p} \right) \Biggr], \\
  m_1^{\rm kin}\kappa (\hat{F}^{\rm IV}_{21}) &= \frac{15}{8\pi}(C_0^\tau - C_1^\tau) \int d\rb (\rho^2 + 2z^2) \rho_n(\rb) \rho_p(\rb), \\
  m_1^{\rm LGSB}(\hat{F}^{\rm IV}_{21}) &= \frac{15}{8\pi}\frac{1}{A^2}
  \int d\rb (\rho^2 + 2z^2) G_{\rm IV}^{\rm LGSB}(\bm{r}), \\
   m_1^{\rm kin}(\hat{F}^{\rm IV}_{22}) &= \frac{5}{2\pi} \frac{\hbar^2}{2m} \frac{NZ}{A^2}
   \Biggl[ Z \langle r_n^2\rangle \left(1 - \sqrt{\frac{5}{4\pi}} \beta_{2n}\right) \nonumber \\
     &\quad + N \langle r_p^2 \rangle \left( 1 - \sqrt{\frac{5}{4\pi}}\beta_{2p}\right) \Biggr], \\
  m_1^{\rm kin}\kappa (\hat{F}^{\rm IV}_{22}) &= \frac{15}{4\pi}(C_0^\tau - C_1^\tau)
  \int d\rb \rho^2 \rho_n(\rb) \rho_p(\rb), \\
  m_1^{\rm LGSB}(\hat{F}^{\rm IV}_{22}) &= \frac{15}{4\pi}\frac{1}{A^2}
  \int d\rb \rho^2 G_{\rm IV}^{\rm LGSB}(\bm{r}), 
  \end{align} \begin{align}
  m_1^{\rm kin}(\hat{F}^{\rm IV}_{30}) &= \frac{63}{16\pi}\frac{\hbar^2}{2m} \frac{1}{A^2}
  \int d\rb (\rho^4 + 4z^4) \nonumber \\ & \times[ Z^2 \rho_n(\rb) + N^2\rho_p(\rb) ], \\
  m_1^{\rm kin}\kappa (\hat{F}^{\rm IV}_{30}) &= \frac{63}{16\pi}(C_0^\tau - C_1^\tau) \int d\rb (\rho^4 + 4z^4) \nonumber\\&\quad\times\rho_n(\rb) \rho_p(\rb), \\
  m_1^{\rm LGSB}(\hat{F}^{\rm IV}_{30}) &= \frac{63}{16\pi} \frac{1}{A^2}
  \int d\rb (\rho^2 + 4z^2) G_{\rm IV}^{\rm LGSB}(\bm{r}), 
    \end{align}\begin{align}
  m_1^{\rm kin}(\hat{F}^{\rm IV}_{31}) &= \frac{21}{32\pi}\frac{\hbar^2}{2m} \frac{1}{A^2}
  \int d\rb (5\rho^4 + 16z^4+16\rho^2z^2)\nonumber\\&\quad\times [ Z^2 \rho_n(\rb) + N^2\rho_p(\rb) ], \\  
  m_1^{\rm kin}\kappa (\hat{F}^{\rm IV}_{31}) &= \frac{21}{32\pi}(C_0^\tau - C_1^\tau) \int d\rb\nonumber\\&\quad\times (5\rho^4 + 16z^4+16\rho^2z^2) \rho_n(\rb) \rho_p(\rb), \\
  m_1^{\rm LGSB}(\hat{F}^{\rm IV}_{31}) &= \frac{21}{32\pi} \frac{1}{A^2}
  \int d\rb \nonumber\\ &\quad \times (5\rho^4 + 16z^4+16\rho^2z^2) G_{\rm IV}^{\rm LGSB}(\bm{r}),
  \end{align}\begin{align} 
  m_1^{\rm kin}(\hat{F}^{\rm IV}_{32}) &= \frac{105}{32\pi}\frac{\hbar^2}{2m} \frac{1}{A^2}
  \int d\rb (\rho^4 + 8\rho^2z^2) \nonumber\\&\quad\times[ Z^2 \rho_n(\rb) + N^2\rho_p(\rb) ], \\
  m_1^{\rm kin}\kappa (\hat{F}^{\rm IV}_{32}) &= \frac{105}{32\pi}(C_0^\tau - C_1^\tau) \int d\rb (\rho^4 + 8\rho^2z^2)\nonumber\\&\quad\times \rho_n(\rb) \rho_p(\rb), \\
  m_1^{\rm LGSB}(\hat{F}^{\rm IV}_{32}) &= \frac{105}{32\pi} \frac{1}{A^2}
  \int d\rb (\rho^4 + 8\rho^2z^2) G_{\rm IV}^{\rm LGSB}(\bm{r}),
    \end{align}\begin{align}
   m_1^{\rm kin}(\hat{F}^{\rm IV}_{33}) &= \frac{315}{32\pi}\frac{\hbar^2}{2m} \frac{1}{A^2}
   \int d\rb\rho^4 [ Z^2 \rho_n(\rb) + N^2\rho_p(\rb) ],\\
   m_1^{\rm kin}\kappa (\hat{F}^{\rm IV}_{33}) &= \frac{315}{32\pi}(C_0^\tau - C_1^\tau) \int d\rb \rho^4  \rho_n(\rb) \rho_p(\rb), \\
   m_1^{\rm LGSB}(\hat{F}^{\rm IV}_{33}) &= \frac{315}{32\pi} \frac{1}{A^2}
   \int d\rb\rho^4 G_{\rm IV}^{\rm LGSB}(\bm{r}),
\end{align}
where
\begin{align}
  G_{\rm IV}^{\rm LGSB}(\bm{r}) &= \sum_{k=0}^1 (C^\tau_k + C^j_k) 
  \left[ Z \rho_n(\bm{r}) + (-1)^{k+1} N\rho_p(\bm{r})\right]^2 \nonumber \\  
  &\quad -(4 \Ct_n^{\Delta\rho} + \Ct_n^\tau) Z^2 |\rhot_n(\bm{r})|^2 \nonumber \\
  &\quad -(4 \Ct_p^{\Delta\rho} + \Ct_p^\tau) N^2 |\rhot_p(\bm{r})|^2.
\end{align}

\bibliographystyle{apsrev4-1}

\end{document}